\documentclass[a4paper,11pt,twoside,bibnote]{cpc-hepnp}
\usepackage{CJK,upgreek,fancyhdr}
\usepackage{multicol}
\usepackage{graphicx}
\usepackage{booktabs}
\usepackage{slashed}
\usepackage{amssymb,bm,mathrsfs,bbm,amscd}
\usepackage[tbtags]{amsmath}
\usepackage{lastpage}
\usepackage{multirow}
\usepackage{textcomp}
\usepackage{epstopdf}
\usepackage{float}
\usepackage{footnote}
\usepackage{subcaption}

\begin{document}

\setlength{\abovecaptionskip}{4pt plus1pt minus1pt}
\setlength{\belowcaptionskip}{4pt plus1pt minus1pt}
\setlength{\abovedisplayskip}{6pt plus1pt minus1pt}
\setlength{\belowdisplayskip}{6pt plus1pt minus1pt}
\addtolength{\thinmuskip}{-1mu}
\addtolength{\medmuskip}{-2mu}
\addtolength{\thickmuskip}{-2mu}
\setlength{\belowrulesep}{0pt}
\setlength{\aboverulesep}{0pt}
\setlength{\arraycolsep}{2pt}

\fancyhead[c]{\small Chinese Physics C~~~Vol. 42, No. 10 (2018) 104001}
\fancyfoot[C]{\small 104001-\thepage}
\footnotetext[0]{Received 16 April 2018}

\title{Event-by-event efficiency fluctuations and efficiency correction for cumulants of superposed multiplicity distributions in relativistic heavy-ion collision experiments\thanks{Supported by the MoST of China 973-Project No.2015CB856901, NSFC (11575069)}}

\author{Shu He, Xiaofeng Luo$^{1)}$\email{xfluo@mail.ccnu.edu.cn}}

\begin{CJK*}{UTF8}{gbsn}
\maketitle
\end{CJK*}

\address{Key Laboratory of Quark \& Lepton Physics (MOE) and Institute of Particle Physics, \\ Central China Normal University, Wuhan 430079, China\\ }

\begin{abstract}
We performed systematic studies on the effects of event-by-event efficiency fluctuations on efficiency correction for cumulant analysis in relativistic heavy-ion collision experiments. Experimentally, particle efficiencies of events measured under different experimental conditions should be different. For fluctuation measurements, the final event-by-event multiplicity distributions should be the superposed distributions of various type of events measured under different conditions. We demonstrate efficiency fluctuation effects using numerical simulation, in which we construct an event ensemble consisting of events with two different efficiencies. By using the mean particle efficiencies, we find that the efficiency corrected cumulants show large deviations from the original inputs when the discrepancy between the two efficiencies is large. We further studied the effects of efficiency fluctuations for the cumulants of net-proton distributions by implementing the UrQMD events of Au+Au collisions at $\sqrt{s_{NN}}=7.7$ GeV in a realistic STAR detector acceptance. We consider the unequal efficiency in two sides of the Time Projection Chamber (TPC), multiplicity dependent efficiency, and the event-by-event variations of the collision vertex position along the longitudinal direction ($V_\mathrm{z}$). When the efficiencies fluctuate dramatically within the studied event sample, the effects of efficiency fluctuations have significant impacts on the efficiency corrections of cumulants with the mean efficiency. We find that this effect can be effectively suppressed by binning the entire event ensemble into various sub-event samples, in which the efficiency variations are relatively small. The final efficiency corrected cumulants can be calculated from the weighted average of the corrected factorial moments of the sub-event samples with the mean efficiency.
\end{abstract}


\begin{pacs}
25.75.Gz, 12.38.Mh, 21.65.Qr     \qquad     {\bf DOI:} 10.1088/1674-1137/42/10/104001
\end{pacs}

\footnotetext[0]{\hspace*{-3mm}\raisebox{0.3ex}{$\scriptstyle\copyright$}2018 Chinese Physical Society and the Institute of High Energy Physics of the Chinese Academy of Sciences and the Institute of Modern Physics of the Chinese Academy of Sciences and IOP Publishing Ltd}%

\vspace{-0.2mm}
\begin{multicols}{2}

\section{Introduction}

The major physics goals of heavy-ion collision experiments are to explore the phase structure of strongly interacting nuclear matter and to study the properties of quark-gluon plasma (QGP)~\cite{Gupta:2011wh,Cleymans:1990nz, Davidson:1991um, Lao:2017skd,Adamczyk:2014ipa, Brachmann:1999xt, Song:2017wtw,Hattori:2016emy}. The QCD phase structure can be displayed in a two-dimensional phase diagram with the temperature \(T\) versus the baryon chemical potential \(\mu_\mathrm{B}\). Lattice QCD calculations show that the transition from a hadronic phase to a QGP phase at zero \(\mu_\mathrm{B}\) is a crossover~\cite{Aoki:2006we} and QCD-based models suggest that at larger \(\mu_\mathrm{B}\), the transition is of the first order~\cite{Schaefer:2011ex, Asakawa:2009aj}. If these model calculations at finite \(\mu_\mathrm{B}\) are correct, there should exist an endpoint of the first-order phase transition line, which is the so-called QCD critical point. Due to the sign problem, the first-principle Lattice QCD calculation becomes very difficult at \(\mu_\mathrm{B}\) $>$ 0~\cite{Endrodi:2011gv}, thus there are large uncertainties in determining the location of the critical point from theoretical calculations~\cite{Stephanov:2004wx, Chen:2015dra, Vovchenko:2015pya, Vovchenko:2016rkn, Fan:2017kym, Fukushima:2014lfa, Karsch:2010ck, BraunMunzinger:2011dn, Gavai:2010zn, Borsanyi:2013hza, Bazavov:2012vg}.  By tuning the collision energy in heavy-ion collisions, QCD matter with various (\(\mu_\mathrm{B}\), \(T\)) can be created to access broad regions of the QCD phase diagram.

One of the most important experimental methods of searching for the critical point is the measurements of the event-by-event fluctuations of conserved quantities, such as the net-baryon ($B$)~\cite{Morita:2014fda, Jiang:2015cnt, Jiang:2015hri}, net-charge number ($Q$)~\cite{Adamczyk:2014fia, Alba:2014eba} and net-strangeness ($S$) number~\cite{Luo:2017faz, Ejiri:2005wq, Stephanov:1999zu, Stephanov:2008qz, Stephanov:2011pb, Bzdak:2016sxg, Luo:2012kja, Friman:2014cua, Kitazawa:2011wh, Kitazawa:2012at, Friman:2011pf, Cheng:2008zh, Luo:2015doi, Aggarwal:2010cw, Zhao:2016djo, Jeon:1999gr, Asakawa:2000wh, Xu:2016mqs, Xu:2016qjd} (And their proxy observables net-kaon~\cite{Adamczyk:2017wsl} and net-proton number fluctuations). The fluctuation observables are sensitive to the correlation length \(\xi{}\), which will diverge near the QCD critical point. The Solenoidal Tracker at the RHIC (STAR) experiment has measured the fluctuation of the net-proton multiplicity (which is a proxy to net-baryon) in Au + Au collisions at \(\sqrt{s_{\text{NN}}}\) = 7.7, 11.5, 14.5, 19.6, 27, 39, 62.4, and 200 GeV, which is taken from the first phase of the RHIC beam energy scan program. The measured forth order net-proton cumulants ration ($\kappa\sigma^2 = C_4/C_2$) of 5\% most central events show a non-monotonic energy (or $\mu_\mathrm{B}$) dependence~\cite{Aggarwal:2010wy, Adamczyk:2013dal, Luo:2015ewa}.

To understand the underlying physics associated with this measurement, we need to perform careful studies on the background contributions, such as the detector efficiency and acceptance effects, volume fluctuations, and other noncritical parameters~\cite{Mukherjee:2016kyu, Mukherjee:2015swa, Nahrgang:2014fza, Bzdak:2012an, Ling:2015yau, Berdnikov:1999ph, He:2016uei, Schuster:2009jv, Sakaida:2014pya}. Owing to the finite detector efficiency, efficiency correction is applied and plays a very important role in cumulant analysis. Generally, the efficiencies are obtained by Monte Carlo (MC) embedding technique~\cite{Abelev:2008ab}.  This allows for the determination of the efficiency, which is the ratio of the matched MC tracks number and the number of input tracks. It contains the effects of both the reconstructed tracking efficiency and acceptance.  In principle, the properties of the efficiency, including fluctuations and acceptance, can be obtained from embedding. However, the embedding sample is only with a limited number of events, and usually, a small fraction of the real data. Thus, with limited statistics of embedding data, it is difficult to capture every detail and property of the entire data sample. The efficiencies are obtained by taking the average within an event sample under different experimental conditions, such as variation of the collision vertex position and the detector performance. The final event-by-event multiplicity distributions should be the superposed distributions of various types of events measured under different experimental conditions. For real data analysis, we usually use the mean efficiency to perform the efficiency corrections for cumulants. This is not problematic if the mean efficiency is used, assuming the efficiency variation is relatively small. However, the problem is that higher order cumulants are sensitive statistics and they are influenced by the bulk properties of events. The average quantity of event ensemble will reduce the details of event-by-event discrepancy, which could be crucial to the cumulants analysis. Experimentally, one needs to implement careful data quality assurance to perform precise measurement studies on efficiencies for data samples.

In our work, we demonstrate the effects of efficiency fluctuations on efficiency correction for cumulants using the average efficiency and provide an effective approach to suppress this effect in future data analysis. This is simulated by injecting particle tracks from UrQMD events into the STAR detector acceptance. The efficiency fluctuations result from the fluctuating collision vertex position and the setting different degree of the asymmetry of the TPC efficiencies. This paper is organized as follow. In Section~2, we will introduce the cumulant observables and the efficiency correction to the cumulants. In Section~3, we will demonstrate the effects of using the mean efficiency with numerical simulation. In Section~4, the effects of using mean efficiency are evaluated using events generated from the UrQMD model with a fast detector simulation. Finally, we will end with a summary.

\section{Cumulants and efficiency correction}\label{cumulants-and-efficiency-correction}

The cumulants of conserved charge are sensitive probes to QCD phase transitions and the QCD critical point, the fourth-order cumulant is proportional to the seventh-order of the correlation length \(C_{4} \propto \xi^{7}\). The cumulants \(C_{1}\) $\sim$ \(C_{4}\) can be defined by moments \(\left\langle N \right\rangle{}\), \(\langle N^{2} \rangle{}\), \(\ldots{}\), \(\langle N^{4} \rangle{}\) as:
\begin{equation}\label{eq:moments-to-cumulants}
\begin{aligned}
    C_{1} & = \left\langle N \right\rangle \\[1mm]
    C_{2} & = \left\langle N^{2} \right\rangle - \left\langle N \right\rangle^{2} \\[1mm]
    C_{3} & = 2\left\langle N \right\rangle^{2} - 3\left\langle N \right\rangle\left\langle N^{2} \right\rangle + \left\langle N^{3} \right\rangle \\[1mm]
    C_{4} & = - 6\left\langle N \right\rangle^{4} + 12\left\langle N \right\rangle^{2}\left\langle N^{2} \right\rangle - 3\left\langle N^{2} \right\rangle^{2} \\[1mm]
          &\quad - 4\left\langle N \right\rangle\left\langle N^{3} \right\rangle + \left\langle N^{4} \right\rangle .
\end{aligned}
\end{equation}
The variance \(\sigma^{2}\), skewness \(S\) and kurtosis
\(\kappa{}\) can be defined as
\[
    \sigma^{2} = C_{2},                                    \quad
    S          = \frac{C_{3}}{{\left( C_{2} \right)}^{3/2}}, \quad
    \kappa     = \frac{C_{4}}{C_{2}^{2}}.
\]
The ratios of the cumulants can be directly compared to the thermodynamic susceptibilities, which can be computed in lattice QCD~\cite{Ejiri:2005wq}.
\[
    S\sigma          = \frac{C_{3}}{C_{2}} = \frac{\chi_{4}}{\chi_{2}}, \quad
    \kappa\sigma^{2} = \frac{C_{4}}{C_{2}} = \frac{\chi_{3}}{\chi_{2}}.
\]
The cumulants measured with detector efficiencies can be recovered by efficiency correction. For example, the mean value can be corrected by:
\[
    \left\langle N \right\rangle           = \frac{\left\langle N \right\rangle_{\text{measure}}}{\epsilon}, \quad
    \left\langle N - \bar{N} \right\rangle = \frac{\left\langle N \right\rangle}{\epsilon_{\text{proton}}} - \frac{\left\langle \bar{N} \right\rangle}{\epsilon_\text{anti-proton}}.
\]
Typically, the efficiencies for proton and anti-proton are different. Thus, we should divide the mean value by corresponding efficiency, respectively.

The efficiency corrections for higher-order cumulants are not as straightforward. One can assume that the response function of detected particles follows a binomial distribution with efficiency parameter \(\epsilon{}\). We can then express the moments in terms of factorial moments and/or the factorial cumulants~\cite{Bzdak:2013pha, Luo:2014rea,Nonaka:2017kko}. The \(r\)th-order factorial moments of a stochastic variable \(N\) can be defined from the expectation of its falling factorial as:
\[
    F_{r} = \left\langle N\left( N - 1 \right)\cdots\left( N - r + 1 \right) \right\rangle.
\]
and the factorial moments can be easily corrected for the binomial efficiency. Suppose the measured factorial moment is \(f_{r}\) with efficiency \(\epsilon{}\), we, therefore, have: (Section~6.1)
\begin{equation}\label{eq:original-to-measured-factorial-moments}
    F_{r} = \frac{f_{r}}{\epsilon^{r}}
\end{equation}
With the efficiency-corrected factorial moments \(F_{1}\) to
\(F_{r}\),

\noindent
we can obtain the moments \(\left\langle N^{r}
\right\rangle{}\)
\begin{equation}\label{eq:factorial-moments-to-moments}
    \left\langle N^{r} \right\rangle = \sum_{i = 0}^{r}{s_{2}\left(r,i \right)F_{r}},
\end{equation}
where the \(s_{2}\) is the Stirling numbers of the second kind. With moments \(\left\langle N^{r} \right\rangle{}\) we can further obtain the cumulants using equation~\eqref{eq:moments-to-cumulants}. In the case where the net-proton cumulant is required, we should introduce a two-dimensional factorial moments and the efficiency correction equation can be written as:
\[
    F_{rs} = \frac{f_{rs}}{\epsilon_\mathrm{p}^{r}\epsilon_{\bar{\mathrm{p}}}^{s}},
\]
where the \(\epsilon_\mathrm{p}^{r}\) is the \(r\)th-order of the proton efficiency and the \(\epsilon_{\bar{\mathrm{p}}}^{s}\) is the \(s\)th-order of anti-proton efficiency. \(f_{rs}\) is defined as:
\begin{multline}
    f_{rs} = \left\langle n_\mathrm{p}\left( n_\mathrm{p} - 1 \right)\cdots\left( n_\mathrm{p} - r + 1 \right) \cdot \right. \\
    \left. n_{\bar{\mathrm{p}}}\left( n_{\bar{\mathrm{p}}} - 1 \right)\cdots\left( n_{\bar{\mathrm{p}}} - s + 1 \right) \right\rangle,
\end{multline}
where \(n_\mathrm{p}\) and \(n_{\bar{\mathrm{p}}}\) are the measured proton and anti-proton numbers, respectively. The conversion from \(F_{rs}\) to \(\left\langle N_{\mathrm{p}}^{r} N_{\bar{\mathrm{p}}}^{s} \right\rangle{}\) is
\[
    \left\langle N_{\mathrm{p}}^{r} N_{\bar{\mathrm{p}}}^{s} \right\rangle = \sum_{i_{1} = 0}^{r}{} \sum_{i_{2} = 0}^{s}{s_{2}\left( r,i_{1} \right)s_{2}\left( s,i_{2} \right) F_{rs} }.
\]
The moments of the net-proton can be expressed as
\begin{align}
    \left\langle N_{\mathrm{p} - \bar{\mathrm{p}}}^{k} \right\rangle  = & \left\langle {\left( N_{\mathrm{p}} - N_{\bar{\mathrm{p}}} \right)}^{k} \right\rangle \nonumber\\
     =& \sum_{i = 0}^{k}{{\left( - 1 \right)}^{i} \binom{k}{i} \left\langle N_{\mathrm{p}}^{k - i}N_{\bar{\mathrm{p}}}^{i}\  \right\rangle}.
\end{align}
It is straightforward to write the net-proton cumulants with equation~\eqref{eq:moments-to-cumulants}.

\subsection{Factorial moments and cumulants of superposed distribution}\label{factorial-moments-of-superposed-distribution}

In this section, we discuss the factorial moments of the superposed distribution. For example, if we have a distribution obtained by the mixture of a Poisson distribution, Gaussian distribution, or some other type of distribution, it is then left to determine the relations between the factorial moments of their superposed distribution and the sub-distributions. The probability density function of the superposed distribution can be expressed as:
\begin{equation}\label{eq:pdf-summation}
    \tilde{P}\left( n \right) = \sum a_{i}P_{i}\left( n \right).
\end{equation}
It describes the probability of detecting \(n\) particles in an event, and the event may be from one of the various types. Therefore, \(\tilde{P}\left( n \right)\) is the summation of the probability of detecting \emph{n} particles from the \(i\)th type \(P_{i}(n)\). \(a_i\) is the weight of \(P_i(n)\).

With \(\tilde{P}(n)\), we can write down the generating function \({\tilde{G}}_{F}\left( s \right)\) for factorial moments
\(F_{r}\)
\begin{align}
    {\tilde{G}}_{F}\left( s \right) & = \sum_{n = 0}^{\infty}{\tilde{P}\left( n \right)s^{n}} \nonumber \\
                                    & = \sum_{n = 0}^{\infty}\sum_{i = 0}^{k}{a_{i}P_{i}\left( n \right)} s^{n}.
\end{align}
We can further write
\begin{equation}\label{eq:avg-of-factorial-moment-generating-function}
    {\tilde{G}}_{F}\left( s \right) = \sum_{i = 0}^{k}a_{i}\sum_{n = 0}^{\infty}{P_{i}\left( n \right)}s^{n} = \sum_{i = 0}^{k}a_{i}G_{F}^{\left( i \right)}.
\end{equation}
We then have the relation between superposed factorial moments given by:
\({\tilde{F}}_{r}\) and \(F_{r}^{\left( i \right)}\)
\begin{align}\label{eq:avg-of-factorial-moment}
    {\tilde{F}}_{r} & = \left.\frac{\partial^{r}}{\partial s^{r}}{\tilde{G}}_{F}\left( s \right) \right|_{s = 1} \nonumber \\
                    & = \sum_{i = 0}^{k}a_{i}\left.\frac{\partial^{r}}{\partial s^{r}}G_{F}^{\left( i \right)}\left( s \right) \right|_{s = 1} \nonumber \\
                    & = \sum_{i = 0}^{k}{a_{i}F_{r}^{\left( i \right)}}.
\end{align}
We find that \({\tilde{F}}_{r}\) is the weighted average of \(F_{r}^{\left( i \right)}\).

However, we will show that we cannot use the average of the cumulants from different types of distributions. First, we note that the superposed cumulants \({\tilde{C}}_{r}\) is not the simple weighted average of \(C_{r}^{\left( i \right)}\). Since the generating function of the cumulants \(K\left( \theta \right)\) can be written as~\cite{Kitazawa:2017ljq}
\begin{equation}
    K\left( \theta \right) = K_{\text{fc}}\left( \mathrm{e}^{\theta} \right).
\end{equation}
The \(K_{\text{fc}}\left( \mathrm{e}^{\theta} \right)\) is the
generating function of the factorial cumulants, and we have:
\begin{equation}
    K_{\text{fc}}\left( \mathrm{e}^{\theta} \right) = \ln{G_{F}\left( \mathrm{e}^{\theta} \right)}.
\end{equation}
Therefore
\begin{equation}
    K\left( \theta \right) = \ln{G_{F}\left( \mathrm{e}^{\theta} \right)}.
\end{equation}
The generating function of the superposed distribution is
\begin{equation}
    \tilde{K}\left( \theta \right) = \ln{{\tilde{G}}_{F}\left( \mathrm{e}^{\theta} \right)} = \ln{\sum_{i = 0}^{k}{a_i G_{F}^{\left( i \right)}\left( \mathrm{e}^{\theta} \right)}}.
\end{equation}
The cumulants \({\tilde{C}}_{r}\) are given by:
\begin{align}
    {\tilde{C}}_{r} & = \left.\frac{\partial^{r}}{\partial\theta^{r}}\tilde{K}\left( \theta \right) \right|_{\theta = 0} \nonumber \\
                    & = \left.\frac{\partial^{r}}{\partial\theta^{r}}\ln{\sum_{i = 0}^{k}{a_i G_{F}^{\left( i \right)}\left( \mathrm{e}^{\theta} \right)}} \right|_{\theta = 0} \nonumber \\
                    & \neq \left.\frac{\partial^{r}}{\partial\theta^{r}}\sum_{i = 0}^{k}{a_i\ln{G_{F}^{\left( i \right)}\left( \mathrm{e}^{\theta} \right)}} \right|_{\theta = 0} \nonumber \\
                    & = \sum_{i = 0}^{k}a_i \left.\frac{\partial^{r}}{\partial\theta^{r}}K^{\left( i \right)}\left( \theta \right) \right|_{\theta = 0} = \sum_{i = 0}^{k} a_i C_{r}^{\left( i \right)}.
\end{align}
Thus, if the individual distributions are different, the cumulant of the superposition of different distributions is not the average of the cumulants of the individual distributions.

Suppose that \(G_F\left(\mathrm{e}^\theta\right)\) is the factorial moment generating function after efficiency correction. Therefore, the superposed cumulant generating function is given as:
\begin{align}
    \tilde{K}(\theta) & = \ln\tilde{G}_F\left(\mathrm{e}^\theta\right) = \ln{\sum_{i=0}^{k}{a_i G^{(i)}_F\left(\mathrm{e}^{\theta}\right)}} \nonumber \\
                      & = \ln{\sum_{i=0}^{k}{a_i G_F\left(\mathrm{e}^{\theta}\right)}}\nonumber \\
                      & = \ln{G_F\left(\mathrm{e}^{\theta}\right)} \nonumber \\
                      & = \sum_{i=0}^{k}{a_i} \ln{ G_F\left(\mathrm{e}^{\theta}\right)} \nonumber \\
                      & = \sum_{i=0}^{k}{a_i} K^{(i)}(\theta).
\end{align}

We find that this relation is only true when all the \(G_F^{(i)}\) are equal. In order words, the average cumulant is only valid for the superposed distributions of the same type.

The statistical error for the superposed cumulants is given by the Delta theorem~\cite{Luo:2014rea, Luo:2011tp}. In our discussion, the detecting efficiency \(\epsilon{}\) is taken as a constant. Therefore, we have:
\begin{align}
    V\left( {\tilde{C}}_{r} \right) & = \sum_{p,q}^{r}{\frac{\partial{\tilde{C}}_{r}}{{\tilde{f}}_{p}}\frac{\partial{\tilde{C}}_{r}}{{\tilde{f}}_{q}}}\text{Cov}\left( {\tilde{f}}_{p},{\tilde{f}}_{q} \right) \nonumber \\
                                    & = \sum_{p,q}^{r}\sum_{i}^{k}{\frac{\partial{\tilde{C}}_{r}}{f_{p}^{\left( i \right)}}\frac{\partial{\tilde{C}}_{r}}{f_{q}^{\left( i \right)}}}
                                    \text{Cov}\left( f_{p}^{\left( i \right)},f_{q}^{\left( i \right)} \right).
\end{align}

\subsection{Efficiency correction for superposed distribution with different efficiencies}\label{efficiency-correction-for-superposed-distribution-with-different-efficiencies}

Experimentally measured multiplicity distribution can be treated as a superposed of distributions with different efficiencies. For simplicity, we assume that the response function of the detected efficiency is a binomial distribution. This is a special case of equation~\eqref{eq:pdf-summation}, where \(p_i (n)\) is given by equation~\eqref{eq:detecting-probability} with a different efficiency \(\epsilon_{i}\) as:
\begin{equation}
    p_i(n) = \sum_{N=n}^{\infty}{P(N)B_{N}(n, \epsilon_i)}
\end{equation}
and the PDF for superposed distribution is:
\begin{equation}
    \tilde{p}(n) = \sum_{i=0}^{k} a_i p_i(n) = \sum_{i=0}^{k}\sum_{N=n}^{\infty}{a_i P(N)B_{N}(n, \epsilon_i)}.
\end{equation}
The generating function of the measured factorial moments for each species of the distribution is given by equation~\eqref{eq:ori-gen-func-to-mea-gen-func} 
\begin{align}\label{CMG2CMG}
    G_{f}^{(i)}\left( s \right) & = \sum_{n=0}^{\infty}{p_i(n)} s^n= \sum_{n=0}^{\infty} \sum_{N=n}^{\infty}{P(N)B_{N}(n, \epsilon_i) s^n} \nonumber \\
                                & = \sum_{N = 0}^{\infty}{P\left( N \right)\left\lbrack 1 + \epsilon_{i}\left( s - 1 \right) \right\rbrack^{N}} \nonumber \\
                                & = \sum_{N = 0}^{\infty}{P\left( N \right){{s'_{i}}^{N}}} = G_{F}(s'_i),
\end{align}
where the $s'_i= 1 + \epsilon_{i}\left( s - 1 \right)$. Their average is given by equation~\eqref{eq:avg-of-factorial-moment-generating-function}, and we have:
\begin{align}
    \tilde{G}_{f}(s)   = \sum_{i}^{k}{a_i G_f^{(i)}(s)}   = \sum_{i}^{k}{a_i G_{F}(s'_i)}.
\end{align}
We then have the relation between the measured factorial moments \(\tilde{f}_r\) and the original factorial moments from each species of event
\begin{align}\label{eq:superposed-factorial-moments-with-efficiencies}
    \tilde{f}_r & =\left.\frac{\partial^r}{\partial s^r}{\tilde{G}_f(s)}\right|_{s=1}\nonumber \\
                & = \sum_{i}^{k}{a_i\left.{\left(\frac{\partial s'_i}{\partial s}\right)}^{r}\frac{\partial^r}{\partial {(s'_i)}^r}G_F\left(s'_i\right)\right|_{s=1}} \nonumber \\
                & = \sum_{i}^{k}{a_i\left.\epsilon_{i}^r\frac{\partial^r}{\partial {(s'_i)}^r}G_F\left(s'_i\right)\right|_{s'_i=1}} \nonumber \\
                & = \sum_{i}^{k}{a_i \epsilon_{i}^r F_r}.
\end{align}

The mean efficiency \(\left\langle\epsilon\right\rangle{}\) should not be used for the superposed distribution. It can be demonstrated in equation~\eqref{eq:superposed-factorial-moments-with-efficiencies} by multiple \(1/\left\langle\epsilon\right\rangle{}\) to both sides, and comparing it to the original superposed factorial moments
\begin{align}\label{eq:diff-of-fact-by-using-mean-eff}
    \frac{\tilde{f}_r}{{\left\langle\epsilon\right\rangle}^r} - {F}_r & = \sum_{i}^{k} a_i \frac{\epsilon_i^r}{{{\left\langle\epsilon\right\rangle}}^r}F_r - {F_r} \nonumber \\
    & = F_r \left(\frac{\sum_{i}^{k} a_i \epsilon_i^r}{{\left\langle\epsilon\right\rangle}^r} - 1 \right) \nonumber \\
    & = F_r \left( \frac{\left\langle\epsilon^r\right\rangle}{{\left\langle\epsilon\right\rangle}^r} - 1\right).
\end{align}
Since \(\epsilon_i\) is fluctuating, the last line is usually not equal to 0. Ideally, in order to obtain factorial moments of the original distribution, we should perform efficiency correction for each types of events, separately:
\begin{equation}\label{eq:com-fact-corr}
F_{r} = \sum_{i = 0}^{k}{a_{i}\frac{f_{r}^{\left( i
\right)}}{\epsilon_{i}^{r}}}.
\end{equation}
There are two methods to obtain the efficiency corrected cumulants for superposed distributions, from distributions with different efficiencies:
\begin{enumerate}
    \item Correct \(f_r^{(i)}\) to \(F_r^{(i)}\) and compute \(C_r^{(i)}\). Then \(\tilde{C}_r\) = \(\sum_i a_i C_r^{(i)}\).
    \item Correct \(f_r^{(i)}\) to \(F_r^{(i)}\), and \(\tilde{F}_r\) = \(\sum_i a_i F_r^{(i)}\). Then compute \(\tilde{C}_r\) from \(\tilde{F}_r\).
\end{enumerate}
If we know the efficiencies of the different event types in the superposed distribution, we should note that after efficiency correction \(F_r^{(1)}\) = \(F_r^{(2)}\) = \(\cdots{}\) = \(F_r^{(k)}\). Therefore, we can demonstrate that methods 1 and 2 are equivalent. However, in Section~{\ref{vz-bin-correction}}, we will find that the efficiencies of each type of events are unknown. In this case, we use the mean \(\epsilon{}\) in each sub-event sample. Thus, the generating functions \(G_F^{(i)}\) in different event sample bins are still not the same after efficiency correction. \(G_F^{(i)}\) is only an approximation to the true \(G_F\). Thus, the correction for superposition of different types of distributions should be performed with the average factorial moments (method 2). If the efficiency variation in each bin is not small, then the weight average of the cumulants will introduce additional uncertainties. We note that this is similar to the case of using the technique of centrality bin width correction (CBWC) to evaluate cumulants in a wide centrality bin to suppress volume fluctuations~\cite{Luo:2013bmi}.

\begin{figure*}
\centering
\includegraphics[width=0.9\textwidth]{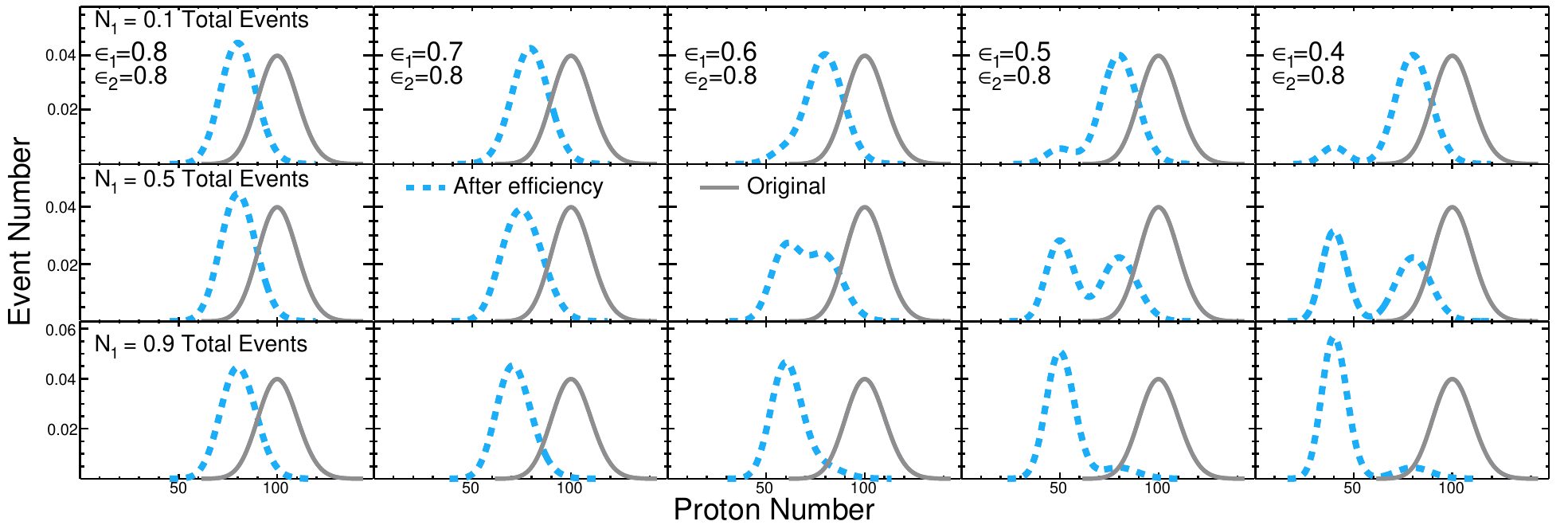}
\figcaption{\label{fig:mc-event-by-event} (color online) Monte Carlo input (Original, gray line) and measured distribution with detector efficiencies (dashed blue line). The statistics of event is 1.0 billion (\(10^{9}\)). In the first, second and third rows, events with efficiency \(\epsilon_{1}\) (i.e., Events of type I) makes up 10\%, 50\% and 90\% of the total event number, respectively. In columns 1 to 5, the \(\epsilon_{1}\) varies from 0.8 to 0.4, while the efficiency of type II is fixed at \(\epsilon_{2}\) = 0.8.}
\end{figure*}

\begin{figure*}
\centering
{\label{fig:mc-average-cumulants}
\includegraphics[width=1.00\columnwidth]{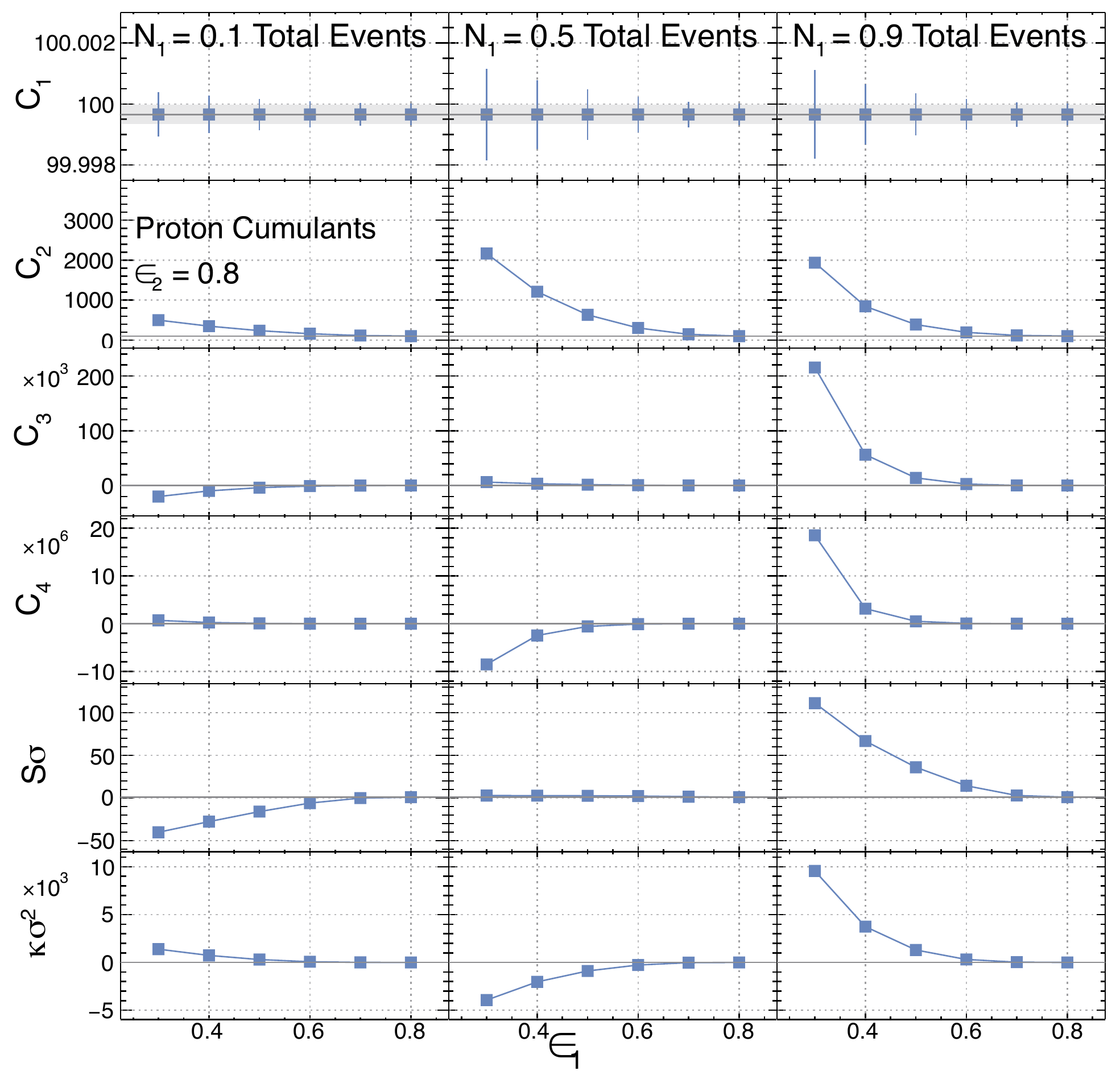}}
{\label{fig:mc-corrected-cumulants}
\includegraphics[width=1.00\columnwidth]{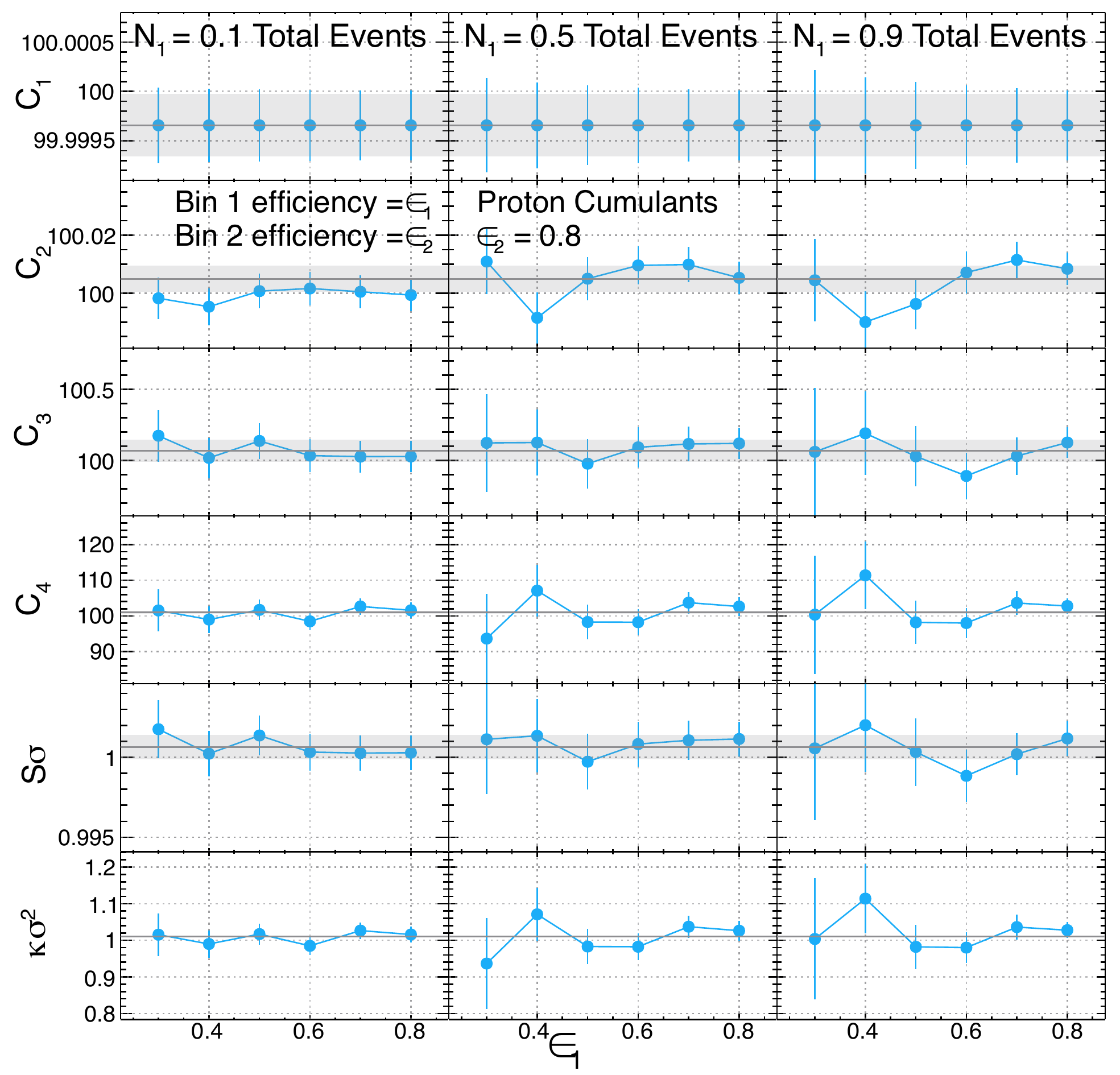}}
\figcaption{\label{fig:mc-cumulants} (color online) Efficiency corrected cumulants and cumulant ratios. For each column, the events of type I makes up 10\%, 50\% and 90\% of the total event number (\(10^{9}\)). The \(\epsilon_{1}\) of the \(x\)-axis represents efficiency event type I\@ and the efficiency of type II is fixed at \(\epsilon_{2}\) = 0.8. \textbf{Square markers} (left): Result corrected with mean efficiency. \textbf{Circle markers} (right): Result corrected independently by \(\epsilon_{1}\) and \(\epsilon_{2}\) (True efficiencies).}
\end{figure*}

\begin{figure*}
\centering
{\label{fig:factorial-moments-and-cumulants-average-at-5-bins}
    \includegraphics[width=0.99\columnwidth]{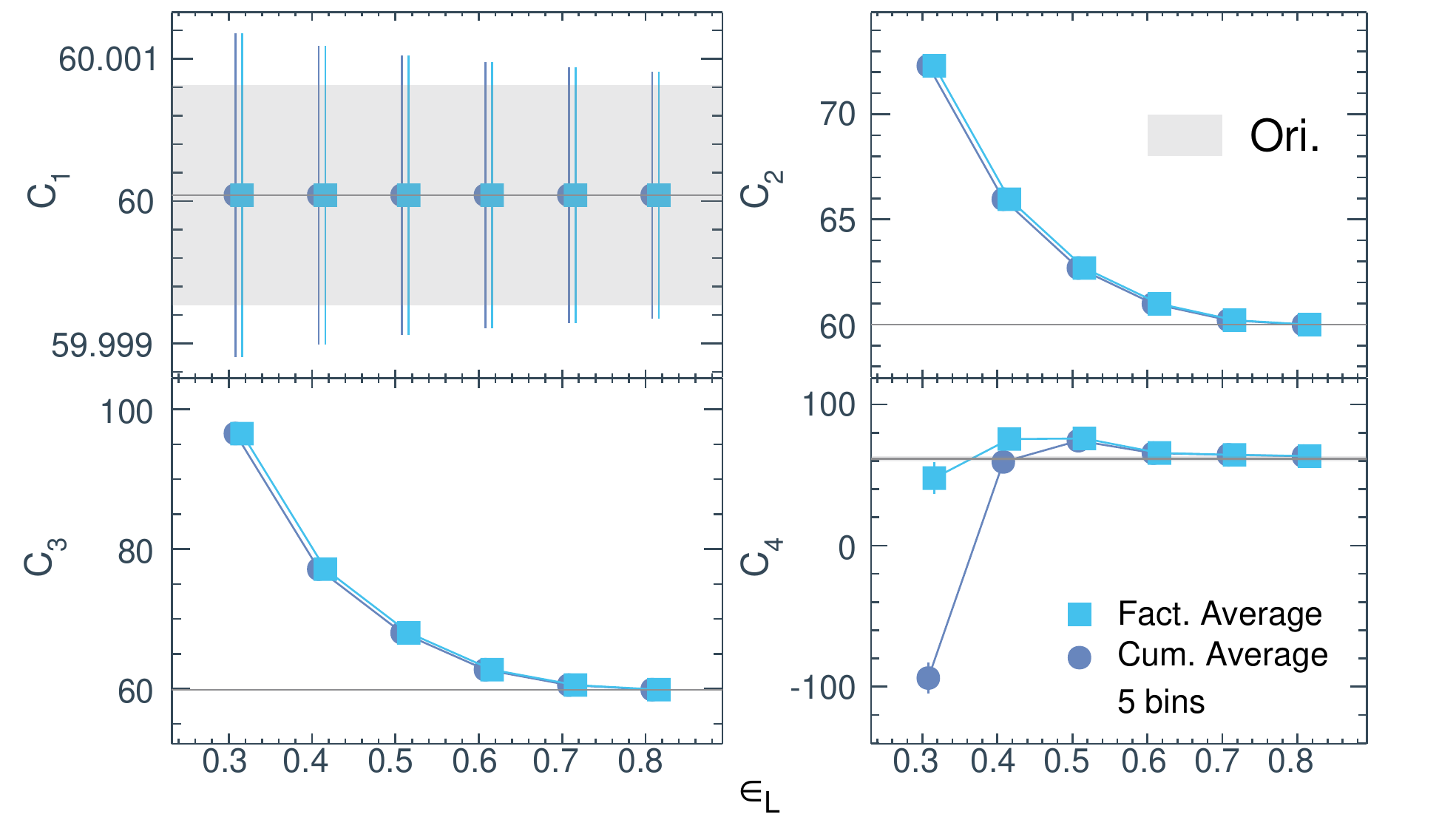}
}
{\label{fig:factorial-moments-and-cumulants-average-at-100-bins}
    \includegraphics[width=0.99\columnwidth]{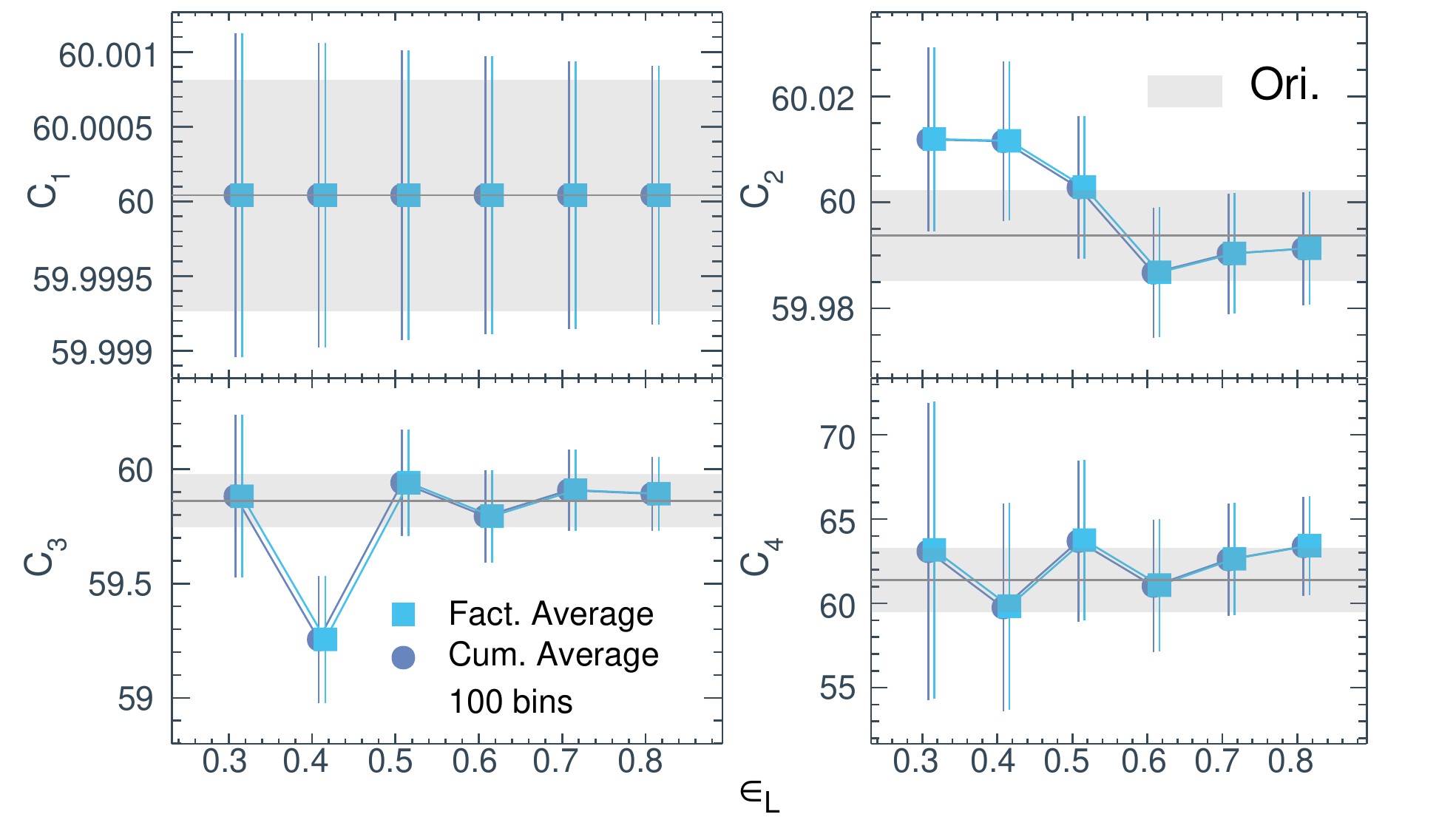}
}
\figcaption{\label{fig:factorial-moments-and-cumulants-average} (color online)  Difference of the factorial moments average and the cumulants average. Simulation with 5 bins (left) and 100 bins (right). The efficiencies of events are selected randomly in the range (\(\epsilon_\mathrm{L}, 0.8\)). With a finer efficiency bin (right), the difference of the two methods become smaller.}
\end{figure*}

\section{Effects of using mean efficiency in efficiency correction for cumulants of multiplicity distributions}\label{problem-of-using-mean-efficiency}

Usually, the efficiency \(\epsilon{}\) in equation~\eqref{eq:original-to-measured-factorial-moments} is obtained from MC embedding. It reflects the net contribution of the detector acceptance, tracking efficiency, and the other effects and it is obtained by taking the average of the entire event ensemble. For most situations, the true efficiency of each event should not fluctuate too far from this mean value. In these cases, the \(\left\langle \epsilon \right\rangle{}\) is a good approximation for correction. But we should be careful when efficiencies of some events dramatically deviate from the $\left<\epsilon\right>$. We can exclude bad events by rejecting events with unusual multiplicity or selecting events within a multiplicity range. It should be noted that with a relative large efficiency shift, the change in \(\left\langle N \right\rangle{}\) can be slight because the binomial distribution is wide. Thus, event selection becomes difficult.

The problem of using mean efficiency to correct cumulants exists in reality. We can consider an extreme example in which an event ensemble mixes two types of distinctive events. One type of event has efficiency \(\epsilon_{1}\), and the other type has efficiency \(\epsilon_{2}\). The mean efficiency eventually determined as the average of the two types of event \(\bar{\epsilon}\). To model this example, we used a Monte Carlo simulation. For each event, the proton number \(N\) we input follows a Poisson distribution with parameter \(\lambda = 100\). In events of type I, the detected proton number \(n\) follows a binomial distribution \(B\left( N,\epsilon_{1} \right)\). In events of type II, the \(n\) follows \(B\left( N,\epsilon_{2} \right)\). The event-by-event proton number distribution from the simulation is shown in Fig.~\ref{fig:mc-event-by-event}.

In Figure.~\ref{fig:mc-event-by-event}, we give the same original input distribution with the number of events ($M$) for each case (the solid grey lines). We then divide the original input events into two sub-event samples, which passes different efficiencies. The two types of events are represented by type I and II with the number of events given as $M_1$ and $M_2$, respectively ($M_1+M_2=M$). The efficiency of type II is fixed at \(\epsilon_{2} = 0.8\). From column 1 to 5, we decrease the efficiency of type I events from 0.8 to 0.4. We find that the distinctive peaks of event-by-event distributions gradually emerged. The event fraction of the total events for the type I sub-event sample are varied as 0.1, 0.5, and 0.9. Then, we perform efficiency correction using the mean efficiency \(\left\langle\epsilon\right\rangle{}\) of each case. Since we can represent \(\epsilon_{1}\) by \(\epsilon_{1} = \left\langle N \right\rangle_{\text{measure}}/\left\langle N \right\rangle_{\text{input}}\) and so is \(\epsilon_{2}\), their average can be written as:
\[
    \left\langle\epsilon\right\rangle = \frac{{M_{1}\left\langle n \right\rangle}_{1} + M_{2}\left\langle n \right\rangle_{2}}{M\left\langle N \right\rangle }.
\]
where \(M_{1}\) is the number of events in type I and \(M_{2}\) is the number of events in type II\@. The measured particle number is denoted as \(\left\langle n \right\rangle{}\), and the input particle number is denoted as \(\left\langle N \right\rangle{}\).

With the measured distributions (blue dashed lines in Fig.~\ref{fig:mc-event-by-event}) and the mean efficiency \(\langle\epsilon\rangle{}\), we can calculate the efficiency corrected factorial moments (equation~\eqref{eq:original-to-measured-factorial-moments}) and the cumulants, which are shown in Fig.~\ref{fig:mc-cumulants} (left) as blue square markers. We then tune the efficiency difference \(\Delta\epsilon{}\) of the two types of events to determine how the efficiency corrected results deviate from the original cumulants (marked as the solid gray lines). We found there is no issue in using the mean efficiency to correct \(C_{1}\), since the results perfectly follow the solid lines (approximately around 100.0 which is the Poisson parameter \(\lambda{}\)) with the change of \(\Delta\epsilon{}\). However, the results start to deviate significantly for \(C_{2}\),  \(C_{3}\) and \(C_{4}\). Obviously, the correction failed even if the \(\Delta\epsilon{}\) is as small as 0.1 (When the event-by-event distribution shows no double peaks in Fig.~\ref{fig:mc-event-by-event}).

Therefore, we know that the event ensemble mixes two types of distinctive event which causes the correction to fail for higher order cumulants. As such, it is necessary to determine whether the results can be improved when we perform corrections on each type. In the following, we independently calculate the cumulants of two types of event and correct them using their own measured efficiencies.
\[
\epsilon_{1} = \frac{{\left\langle N
\right\rangle}_{\text{measure}}^{\left( 1 \right)}}{{\left\langle
N \right\rangle}_{\text{input}}^{\left( 1 \right)}}, \quad
\epsilon_{2} = \frac{{\left\langle N
\right\rangle}_{\text{measure}}^{\left( 2 \right)}}{{\left\langle
N \right\rangle}_{\text{input}}^{\left( 2 \right)}}.
\]
Therefore, we need to determine how to combine the corrected result of different types of events. The simulation of two types of events is the simplest case. Let's consider the measured distribution from a combination of \(K\) types of events. We can find in Section~{\ref{factorial-moments-of-superposed-distribution}} that Equation~\eqref{eq:avg-of-factorial-moment-generating-function} shows that the factorial moments \(f_{r}\) of the superposed distribution is the weighted average of the factorial moments \(f_{r}^{\left(i\right)}\) of each type. Therefore, the efficiency corrected factorial moments of the superposed distribution is:
\begin{equation}\label{eq:efficiency-bin-average-for-corrected-factorial-moments}
F_{r} = \sum_{i = 1}^{k}{a_{i}\frac{f_{r}^{\left( i
\right)}}{\epsilon_{\left( i \right)}^{k}}}.
\end{equation}
The results are shown in Fig.~\ref{fig:mc-cumulants} (right). As expected, the efficiency corrected cumulants follow the input values perfectly in all three cases.

As we have discussed in Section~{\ref{efficiency-correction-for-superposed-distribution-with-different-efficiencies}}, the cumulants of the superposition of different distributions (i.e.,\ distribution corrected using mean efficiencies instead of true efficiencies) should be calculated from averaged factorial moments. The average of the cumulants will introduce additional deviation. We show the difference of the factorial moments average and the cumulants average in Fig.~\ref{fig:factorial-moments-and-cumulants-average}.  In this figure, the results of a numerical simulation with 100M events is presented. Instead of using 2 different efficiencies in the previous simulation, each event is randomly assigned an efficiency number, which is uniformly distributed in the range (\(\epsilon_\mathrm{L}\), 0.8). To perform efficiency correction, the entire event sample is divided into sub-event samples with equal efficiency intervals between (\(\epsilon_\mathrm{L}\), 0.8). The efficiency correction for each sub-event sample is performed using the methods of factorial moments average and the cumulants average, respectively. We can infer from the left panel of Fig.~\ref{fig:factorial-moments-and-cumulants-average} that with 5 efficiency bins (larger efficiency variations in each bin), the efficiency corrected results fail to reproduce the higher-order input cumulants for both methods. However, for the fourth-order cumulant ($C_4$), the average of the factorial moments is closer to the original, and the results of the cumulants average method exhibit large deviations. In order to reproduce the original cumulants, we have to reduce the efficiency variation and use more efficiency bins (with 100 bins in the right panel of Fig.~\ref{fig:factorial-moments-and-cumulants-average}). For finer efficiency bins, the factorial moment generating function \(G_F^{(i)}\) with a mean efficiency becomes closer to its true value \(G_F\). Moreover, the additional uncertainties of the cumulant average are much smaller.

In conclusion of this section, we demonstrate the effects of using the mean efficiency in the efficiency correction for cumulants of multiple distributions. If the efficiency variation within the event sample is large, it is incorrect to use the mean efficiency to perform the efficiency corrections. To perform precise and reliable efficiency correction, one has carefully bin the events into various sub-event samples, in which the efficiency variation is relatively small.

\section{{UrQMD} Model Simulation with STAR Detector Acceptance }\label{event-model-simulation}

\begin{figure}
\begin{minipage}[t]{0.5\textwidth}
\centering
\includegraphics[width=0.9\textwidth]{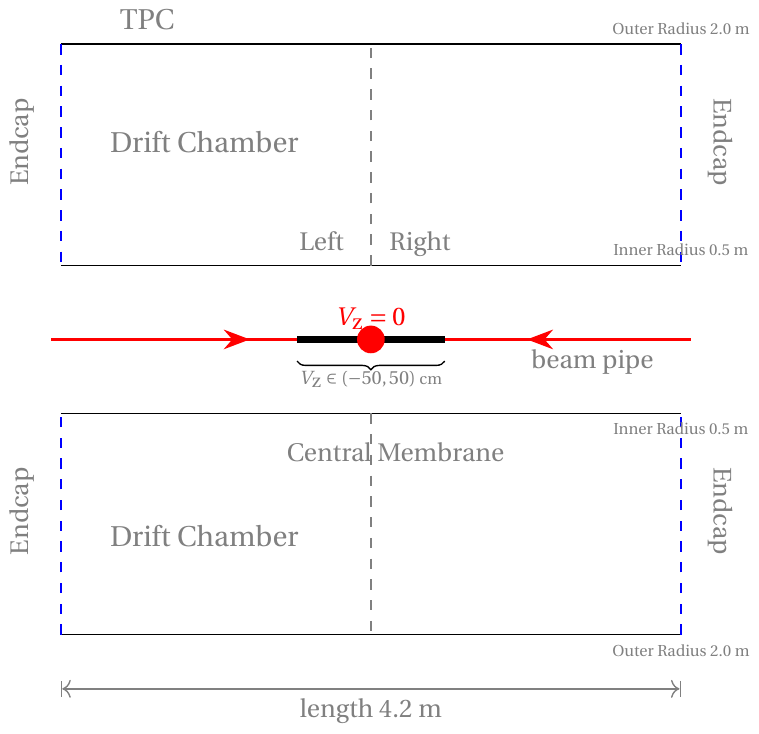}
\figcaption{\label{fig:TPC-structure} (color online) A sketch of the STAR TPC\@.}
\end{minipage}
\begin{minipage}[t]{0.5\textwidth}
\centering
\raisebox{1.8cm}{\includegraphics[width=0.9\textwidth]{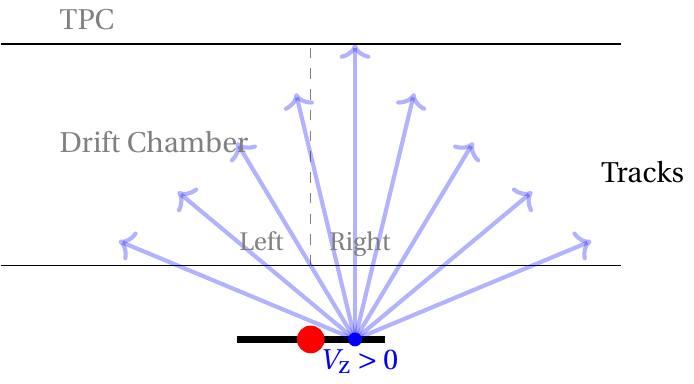}}
\figcaption{\label{fig:geometry-sketch} (color online) Geometry sketch in UrQMD simulation. Each event from UrQMD has been assigned a random \(V_\mathrm{z}\).}
\end{minipage}
\end{figure}

In this section, we will examine whether or not the failed correction in the last section can occur in real experiments. In the STAR experiment, particle identification and track reconstruction are performed with a time projection chamber~\cite{Leo:1987kd} (TPC). The major structure of the TPC is a cylinder drift chamber with a high voltage electrode in the center. The two endcaps of the drift chamber are covered with thin-gap, multiwire proportional chambers (MWPC). The particles that pass through the TPC will experience energy loss due to the ionization of the drifting electrons. By measuring the drift time and the number of electrons collected at the endcaps, we can build the track of arrival particles and calculate their energy loss \(\mathrm{d}E/\mathrm{d}x\). As shown in Fig.~\ref{fig:TPC-structure}, the voltage electrode in the center (Central Membrane) divides the drift chamber into two subparts. Usually, the working conditions of the west and east side of the TPC endcaps are not essentially the same, which can result in different detection efficiencies for the west and east TPC\@.

The \(z\)-coordination of the primary vertex is described by an important event parameter \(V_\mathrm{z}\). \(V_\mathrm{z}\) = 0 indicates that the primary vertex is located at the longitudinal center of the TPC\@. In the simulation, a positive \(V_\mathrm{z}\) indicates that the primary vertex is located to the right. We also set a flat distribution of \(V_\mathrm{z}\) within the range (-50 cm, 50 cm), which is a similar case to RHIC BES at low energies. In our discussion, \(V_\mathrm{z}\) distributions are important because the efficiencies are unequal in the left and the right parts of the chamber. Since the particles from events with \(V_\mathrm{z}\) $<$ 0 are more likely to travel into the left chamber, the mean efficiency of events with \(V_\mathrm{z}\) $< $0 become different from that of events with \(V_\mathrm{z}\) $>$ 0. Obviously, we will arrive at the situation which has been discussed in Section~{\ref{problem-of-using-mean-efficiency}}. In fact, the detecting efficiencies have been observed to change with \(V_\mathrm{z}\) in real experiments.

\end{multicols}

\begin{multicols}{2}
\subsection{Efficiency generation in UrQMD model}\label{efficiency-generation-in-urqmd-model}

To investigate the effect of fluctuation of \(V_\mathrm{z}\), we performed a fast simulation using the UrQMD model. UrQMD is a transport model that can simulate nucleus-nucleus collision events~\cite{Bass:1998ca}. The main idea is to give each UrQMD event a random \(V_\mathrm{z}\) in the range -50 $<$ \(V_\mathrm{z}\) $<$ 50 cm. Then, we can assign efficiency to each particle base on its $\eta$ and the $V_\mathrm{z}$ of the event.

We then simplify the geometry of the TPC into a plane to emphasize the effects of interest. Since we can judge which part of the drift chamber the particle will travel into by its pseudo-rapidity \(\eta{}\), the tracks' azimuthal angle (\(\phi{}\)) can be omitted from our analysis. Therefore, our realm of interest can be represented as shown in Fig.~\ref{fig:geometry-sketch}. In this figure, the \(V_\mathrm{z}\) of an event is randomly assigned. The angle (\(\theta{}\)) between a track and the beam pipeline can be derived from the pseudo-rapidity \(\eta{}\).

Owing to the magnetic field in TPC, particles have a helix trajectory when passing through the chamber. However, we can simplify this motion as a straight line because we are only concerned with the part of the chamber where the track will occur. It should be noted that tracks sometimes go through the central membrane. For the sake of clarity and simplicity, we suppose that the entire track is in the left/right part of the chamber if the end of the track is in the left/right part of the chamber. Finally, we can assign the detection efficiencies \(\epsilon_\mathrm{L}\) to tracks in the left chamber and \(\epsilon_\mathrm{R}\) to tracks in the right chamber.

\begin{figure*}
\centering
\includegraphics[width=0.80\textwidth]{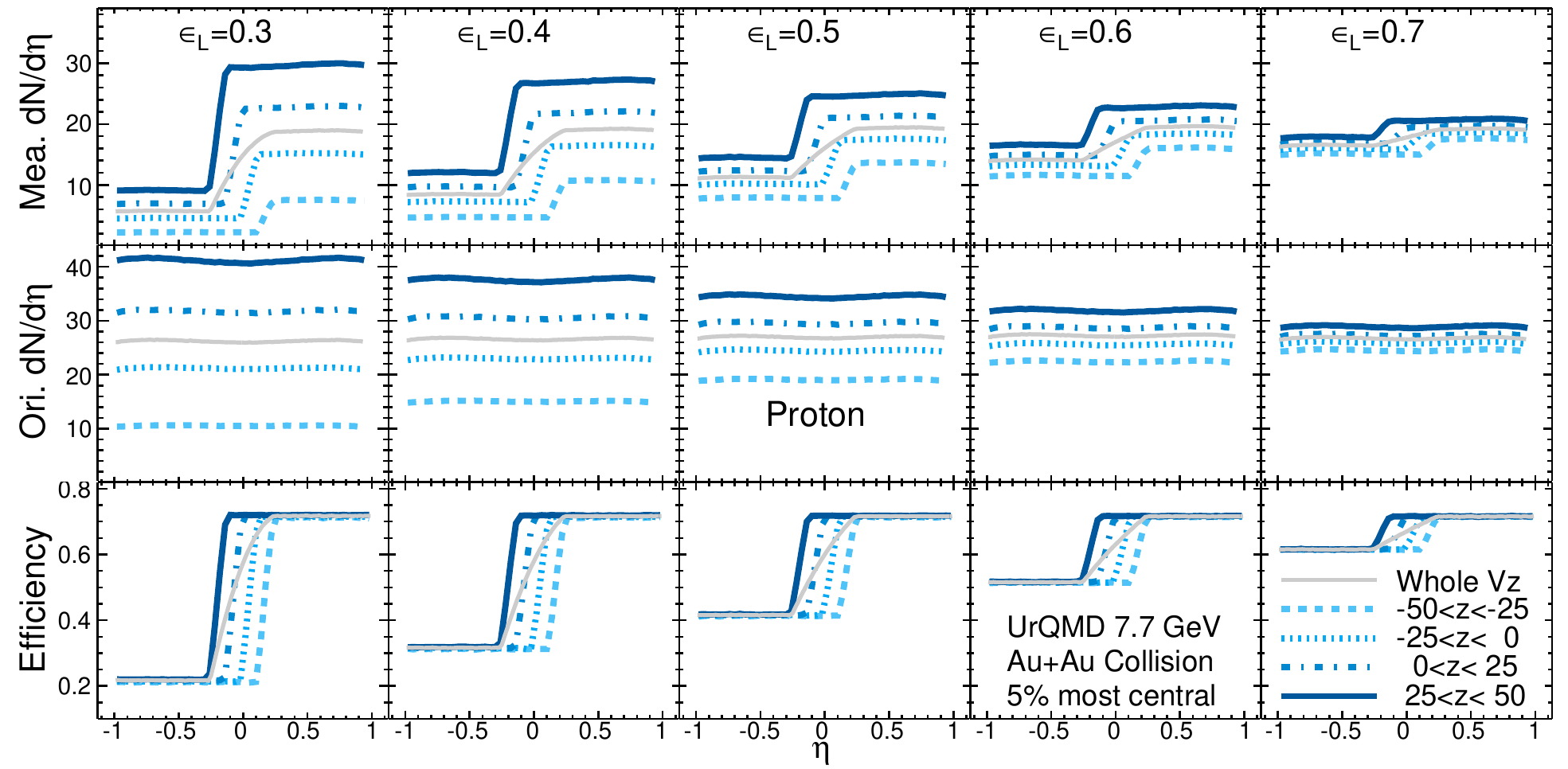}
\figcaption{\label{fig:dN/deta} (color online) \(\mathrm{d}N/\mathrm{d}\eta{}\) distribution (row 1: measured, row 2: original input) and the measurement efficiency (row 3). From left to right columns, the detecting efficiency of tracks in the left part of the drift chamber arising from 0.3 to 0.7, while the efficiency of tracks in the right part of the chamber is fixed at 0.8. The \(\eta{}\) dependence of the detecting efficiency can be represented by the ratio of the first and the second row.}
\end{figure*}

In addition to the effect of \(V_\mathrm{z}\) fluctuation, the detecting efficiency is affected by the total multiplicity of charged particles. The multiplicity of charged particles is usually used as the reference to determine the centrality. This implies that  the detecting efficiency must be different from the central to peripheral collisions. Thus, we introduce a multiplicity dependence efficiency. The relation between total multiplicity and efficiency can be expressed as:
\begin{equation}\label{eq:refmult-dependent-efficiency}
\epsilon = \epsilon_{0} - K_\mathrm{R}N_{\text{mul}},
\end{equation}
where the \(\epsilon_{0}\) and the \(K_\mathrm{R}\) are constant, and the \(N_{\text{mul}}\) is the total multiplicity of charged particles within $|\eta|<1$. The minus sign before \(K_\mathrm{R}\) indicates the detecting efficiency decreases with increasing total multiplicity. This effect can be introduced in the simulation by simply reducing the \(\epsilon_\mathrm{L}\) and \(\epsilon_\mathrm{R}\) by the minus of the factor \(K_\mathrm{R}N_\mathrm{mul}\). For simplicity and clarity, we set the detecting efficiencies has no dependence of $p_\mathrm{T}$ and the efficiencies of the $p(\bar{p})$ are the same. Since we are interested in the efficiency fluctuations effects on net-proton cumulants, we did not apply efficiencies to pions and kaons.

\begin{figure*}
\begin{minipage}[t]{0.48\textwidth}
\centering
\includegraphics[width=0.99\textwidth]{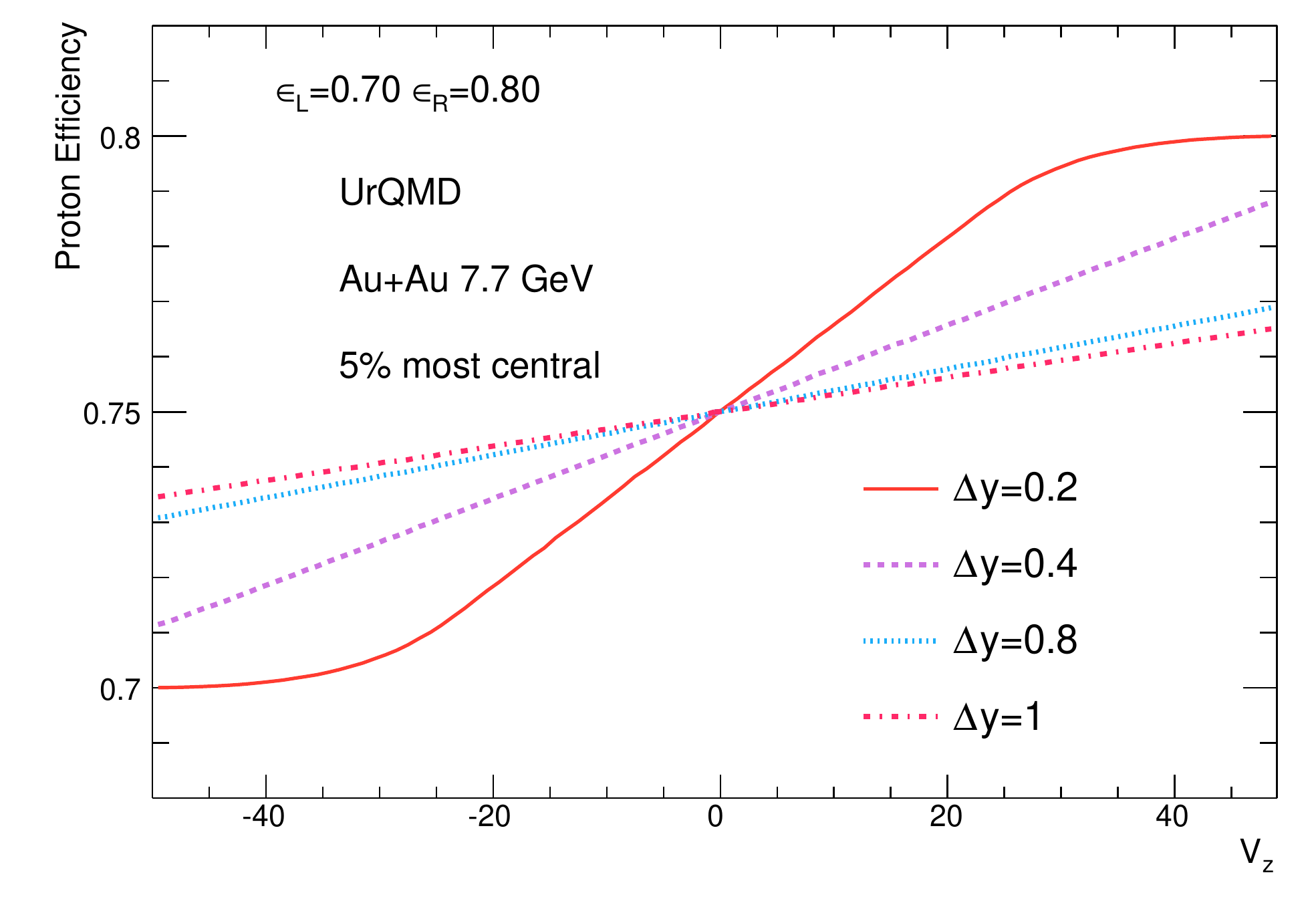}
\figcaption{\label{fig:vz-dy-dependent-efficiency} (color online) \(V_\mathrm{z}\) dependence of detecting efficiency within various rapidity coverage \(\Delta y\). The wider \(\Delta y\) coverage corresponds to a smoother slope in the figure.}
\end{minipage}
\begin{minipage}[t]{0.48\textwidth}
\centering
\includegraphics[width=1.02\textwidth]{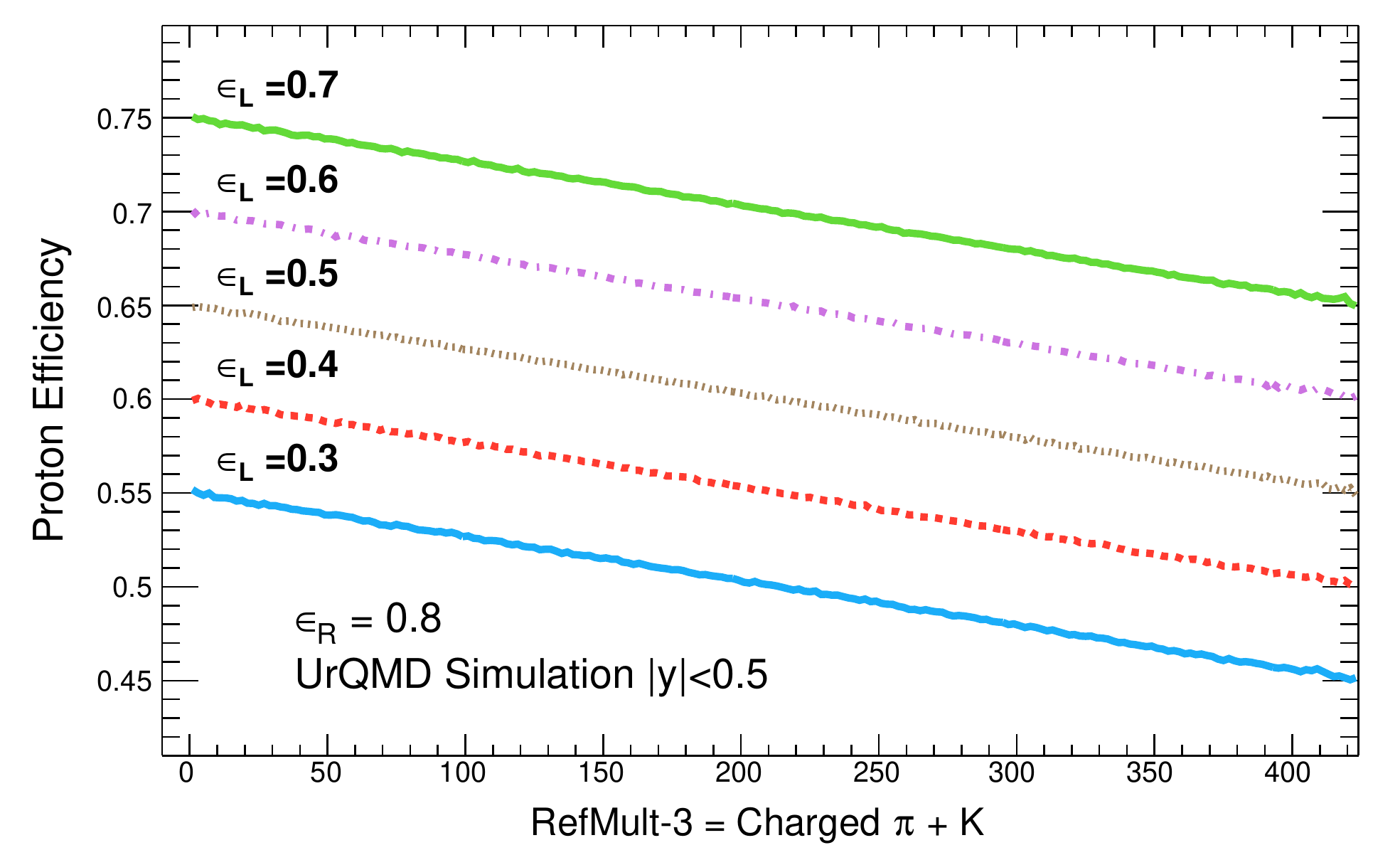}
\figcaption{\label{fig:refmult-depdendent-efficiency} (color online)  Multiplicity dependence of detecting efficiency. The reference multiplicity in the \emph{x}-axis represents the multiplicity of charged \(\pi{}\) and \(K\) meson within \(\left| \eta \right|\) < 1. The different lines \(\epsilon_\mathrm{L}\) in the figure represent various \(\epsilon_{0}\) in equation~\eqref{eq:refmult-dependent-efficiency}.}
\end{minipage}
\end{figure*}

\subsection{Results}\label{result}

In this work, we calculate the efficiency-corrected cumulants of The net-proton distributions in Au + Au collisions at \(\sqrt{s_{\text{NN}}}\) = 7.7 GeV from UrQMD and select 0--5\% most central events from the dataset. The statistics of selecting events is 2.0 million. Collision centrality is determined by the charged particles within $|\eta|<1$ by excluding the proton and anti-protons. In Fig.~\ref{fig:dN/deta}, we show the proton \(\mathrm{d}N/\mathrm{d}\eta{}\) distribution for measured data (with \(V_\mathrm{z}\) fluctuation and detecting efficiencies) and the original UrQMD data within a pseudo-rapidity coverage \(\left| \eta \right|\) $<$ 1.0, which is the same coverage as the TPC of the STAR detector. The lines of different colors in Fig.~\ref{fig:dN/deta} represents the \(\mathrm{d}N/\mathrm{d}\eta{}\) distributions of protons within various \(V_\mathrm{z}\) ranges. In the columns from left to right, we gradually increase the efficiency of the tracks in the left part of the chamber (corresponding to \(\eta{}\) $< $0 from \(\epsilon_\mathrm{L}\) = 0.3 to \(\epsilon_\mathrm{L}\) = 0.7, while we fix the efficiency in the right part of the chamber at \(\epsilon_\mathrm{R}\) = 0.8). Thus, we can evaluate the efficiency fluctuation effects when we enlarge or reduce the difference of \(\epsilon_\mathrm{L}\) and \(\epsilon_\mathrm{R}\).

We can learn from Fig.~\ref{fig:dN/deta} that the proton \(\mathrm{d}N/\mathrm{d}\eta{}\) distributions are asymmetry in the positive and negative \(\eta{}\) regions, while the distributions of the original input are flat. The detecting efficiencies of particles can be expressed as the ratio of the first and the second row. We found that when we narrow the \(V_\mathrm{z}\) bin width, the slope from negative \(\eta{}\) to positive \(\eta{}\) becomes steeper. On the contrary, a wider \(V_\mathrm{z}\) bin resulted in a smaller slope. This implies that a wider \(V_\mathrm{z}\) bin mixes more distinctive events.

The relations between rapidity coverage, \(V_\mathrm{z}\) and the proton efficiency are shown in Fig.~\ref{fig:vz-dy-dependent-efficiency}. In this figure, the mean proton efficiency within various rapidity coverage \(\Delta y\) is plotted as a function of \(V_\mathrm{z}\). The efficiencies for \(V_\mathrm{z}\) $>$ 0 are larger than those for \(V_\mathrm{z}\) $<$ 0, which is consistent with our setting. When the \(\Delta y\) is small, particles are more likely to concentrate in the left or right chamber. Therefore, the slope from \(V_\mathrm{z}\) $<$ 0 to \(V_\mathrm{z}\) $>$ 0 is steeper. When \(\Delta y\) is larger; particles are more dispersed into different chamber parts, which leads to a smooth transition from \(V_\mathrm{z}\) $< $0 to \(V_\mathrm{z}\) $>$ 0.

The effect of the total multiplicity on the detecting efficiency is shown in Fig.~\ref{fig:refmult-depdendent-efficiency}, where the proton mean efficiency is plotted as a function of the multiplicity of charged \(\uppi{}\) and \(K\). Experimentally, instead of using wider centrality bins to calculate the cumulants, the net-proton cumulants are calculated in individual reference multiplicity bins to reduce the volume fluctuation which arises from the uncertainty of the collision geometry. This is the so-called centrality bin width correction (CBWC) technique~\cite{Luo:2013bmi}. The reference multiplicity is equal to the multiplicity of charged \(\pi{}\) and \(K\) within \(\left| \eta \right|\) $<$ 1. From the previous discussion, the mean efficiency should be different across centralities and reference multiplicity bins.

\begin{figure}[H]
\centering
\includegraphics[width=0.99\columnwidth]{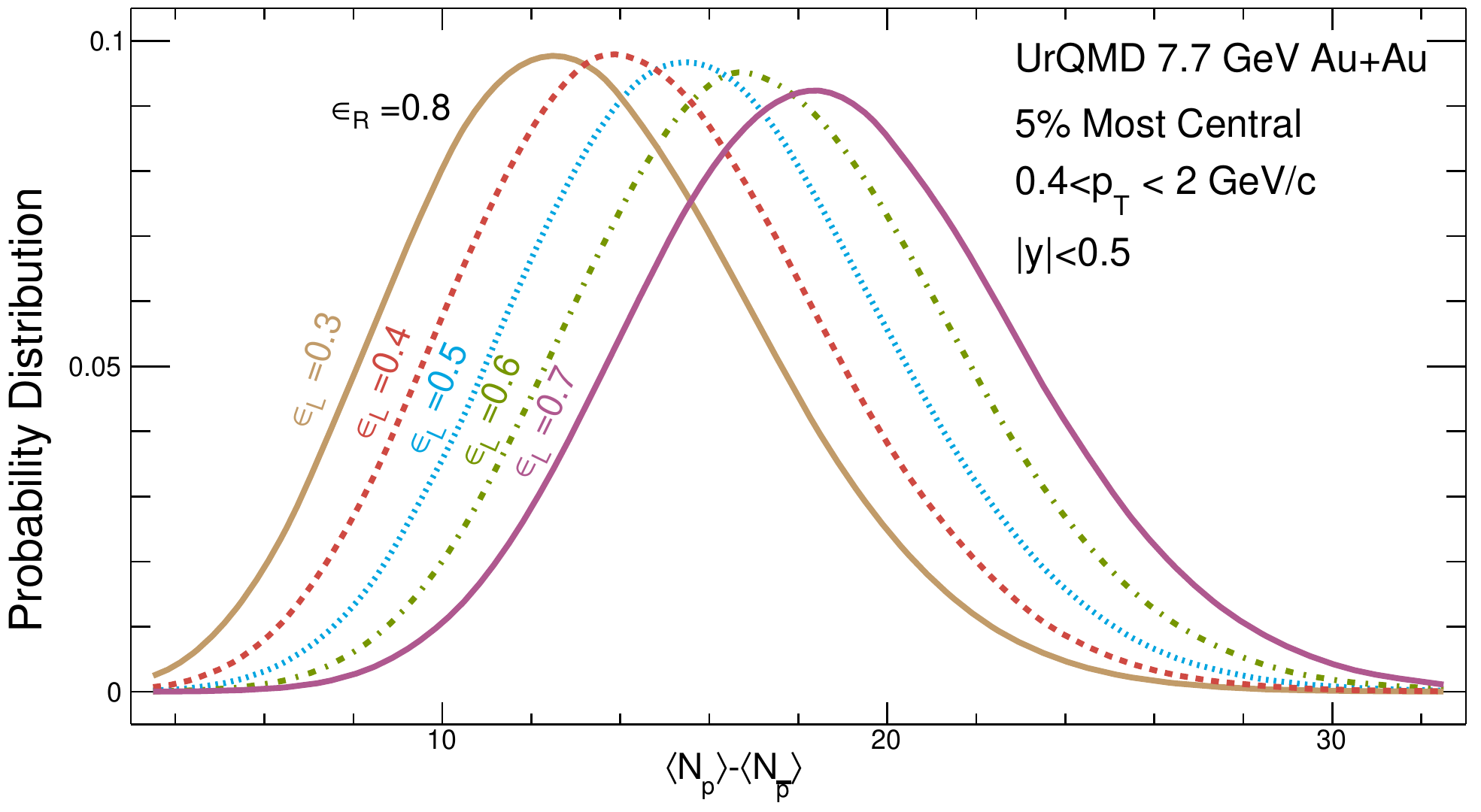}
\figcaption{\label{fig:urqmd-event-by-event-distribution} (color online)  Event-by-event distribution of net-proton. From left to right, the detecting efficiency of tracks in the left part of the drift chamber that arise from 0.3 to 0.7, while the efficiency of tracks in the right part of the chamber is fixed at 0.8.}
\end{figure}

In Fig.~\ref{fig:dN/deta}--\ref{fig:refmult-depdendent-efficiency}, we obtained the proton mean efficiency via simulation and thus we can perform the efficiency correction on the net-proton cumulants. We first examine the event-by-event distributions of the measured net-proton number via simulation. In Fig.~\ref{fig:urqmd-event-by-event-distribution}, we found that the shapes of the distributions show no significant change when we enlarge the difference between \(\epsilon_\mathrm{L}\) and \(\epsilon_\mathrm{R}\). This result is due to the continuous distributions of \(V_\mathrm{z}\). Moreover, the event-by-event distribution of the mean net-proton number is the superposition of events within the whole \(V_\mathrm{z}\) range.

\begin{figure*}
\centering
\includegraphics[width=0.9\textwidth]{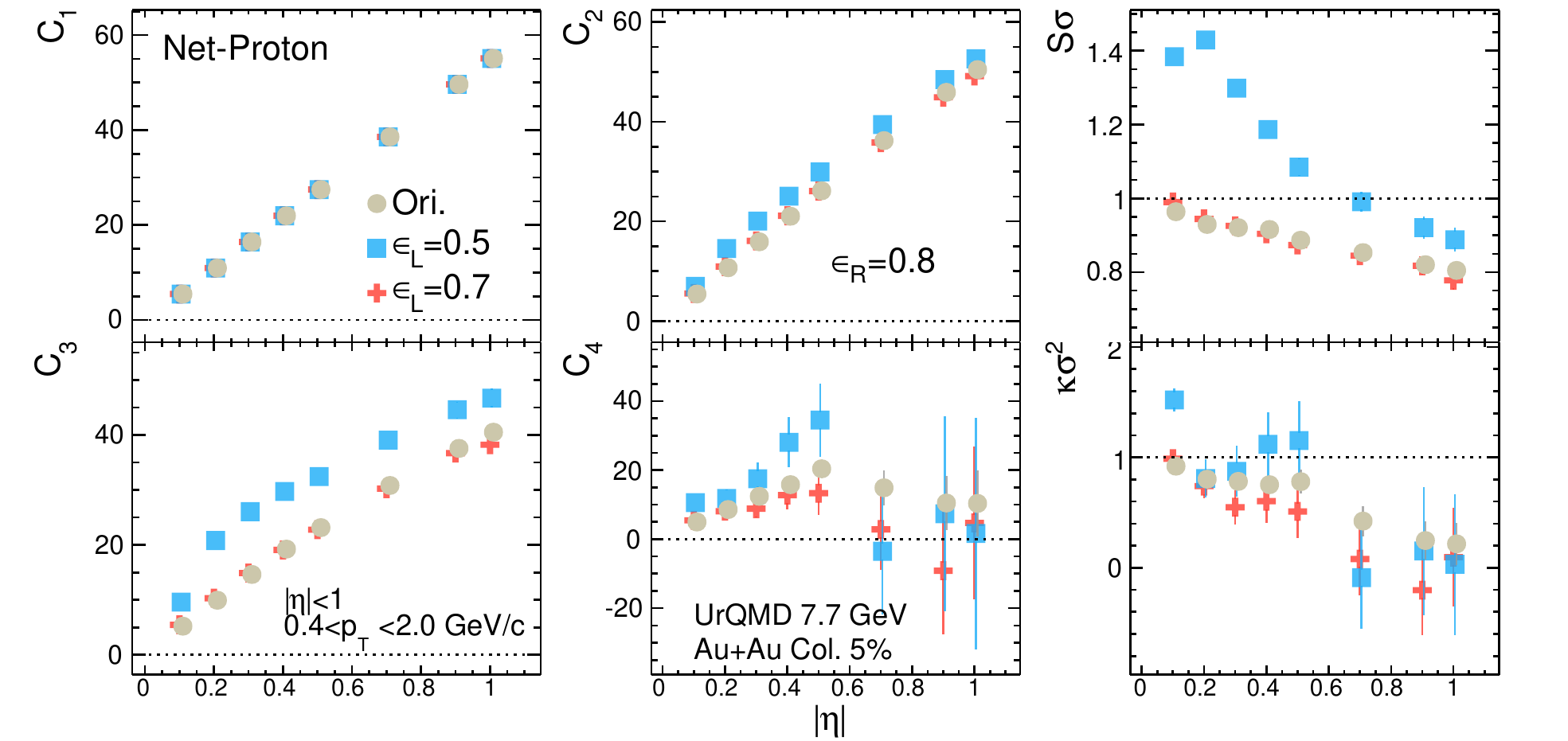}
\figcaption{\label{fig:urqmd-cumulants-to-deta} (color online)  Pseudo-rapidity dependence of efficiency corrected cumulants \(C_1\) to \(C_4\) (and their ratios \(S\sigma = C_3/C_2\), \(\kappa\sigma^2 = C_4/C_2\)) of net-proton distributions. The mean efficiencies \(\left\langle\epsilon_\mathrm{p}\right\rangle\), \(\left\langle\epsilon_\mathrm{\bar{p}}\right\rangle\) are used in the efficiency corrections.}
\end{figure*}

\begin{figure*}
\centering
\includegraphics[width=0.9\textwidth]{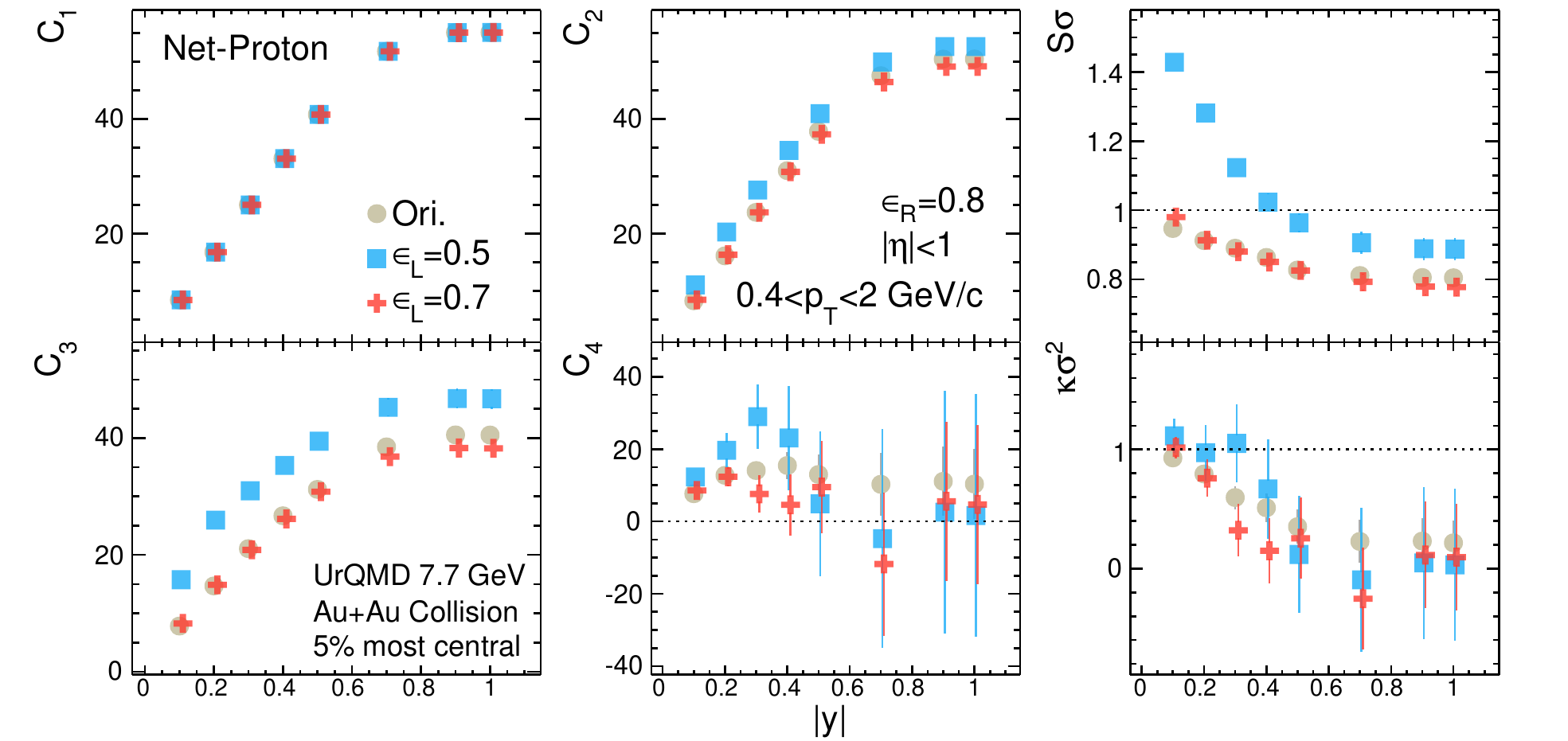}
\figcaption{\label{fig:urqmd-cumulants-to-dy} (color online)  Rapidity dependence of efficiency-corrected cumulants \(C_1\) to \(C_4\) (and their ratios \(S\sigma = C_3/C_2\), \(\kappa\sigma^2 = C_4/C_2\)) of net-proton distributions. The mean efficiencies \(\left\langle\epsilon_\mathrm{p}\right\rangle\), \(\left\langle\epsilon_\mathrm{\bar{p}}\right\rangle\) are used in the efficiency corrections.}
\end{figure*}

\begin{figure*}
\centering
\includegraphics[width=0.9\textwidth]{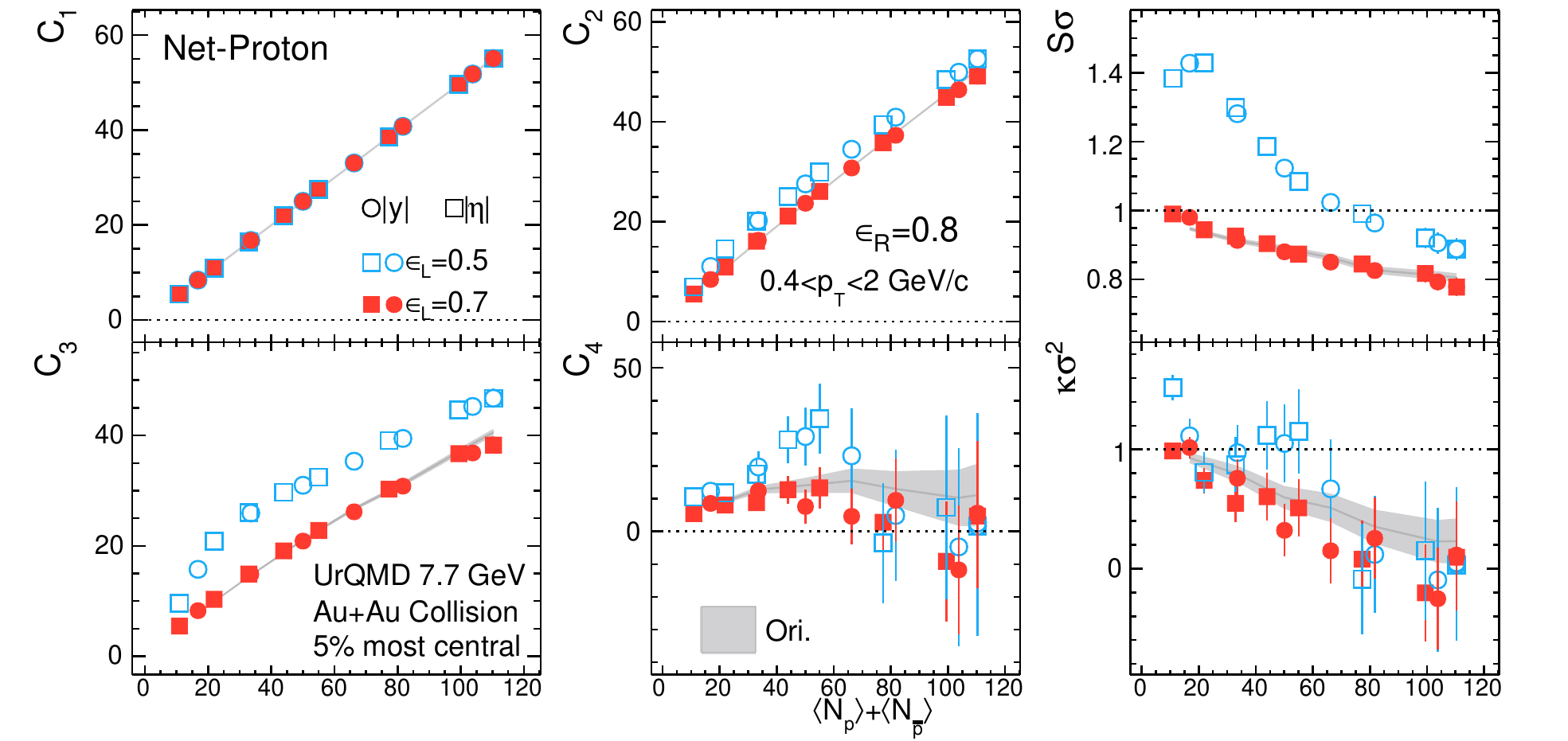}
\figcaption{\label{fig:urqmd-cumulants-to-np} (color online)  Net-proton cumulants and cumulants ratios as functions of mean total-proton number \(\left\langle N_\mathrm{p} \right\rangle + \left\langle N_\mathrm{\bar{p}} \right\rangle\).}
\end{figure*}

We show the efficiency-corrected net-proton cumulants \(C_{1}\) to \(C_{4}\) (and their ratios \(S\sigma{}\) = \(C_{3}/C_{2}\), \(\kappa\sigma^{2}\) = \(C_{4}/C_{2}\)) within various pseudo-rapidity and rapidity coverages in Fig.~{\ref{fig:urqmd-cumulants-to-deta}}--{\ref{fig:urqmd-cumulants-to-dy}}. The mean efficiencies of the proton and anti-protons are used in the efficiency correction. As we suppose in our previous work, the cumulants in various rapidity or pseudo-rapidity acceptance have \(\left\langle N_{\mathrm{p}} \right\rangle{}\) scaling behavior (or \(\left\langle N_{\mathrm{p}} \right\rangle{}\)+\(\left\langle N_{\bar{\mathrm{p}}} \right\rangle{}\) scaling)~\cite{He:2017zpg}. We also plot the cumulants and their ratios as functions of the mean total-proton number \(\left\langle N_{\mathrm{p}} \right\rangle{}\) + \(\left\langle N_{\bar{\mathrm{p}}} \right\rangle{}\) in Fig.~{\ref{fig:urqmd-cumulants-to-np}}. In this case, the cumulants within various \(\Delta\eta{}\) or \(\Delta y\) acceptance show a unified trend to the mean net-proton number (or total-proton number). The original results computed from the original input without detecting efficiency are shown as a solid gray line in the figure. The measured cumulants are represented by colored markers. We found that the efficiency-corrected cumulants coincide with the original results when the efficiencies to the left and right parts of the chamber are close to each other. However, when the difference between \(\epsilon_\mathrm{L}\) and \(\epsilon_\mathrm{R}\) is large (i.e.,\ the case \(\epsilon_\mathrm{L}\) = 0.5, \(\epsilon_\mathrm{R}\) = 0.8), the efficiency correction failed for the higher-order cumulants as was demonstrated in Section~{\ref{problem-of-using-mean-efficiency}}, and the deviations grow rapidly with the difference.

The simulation confirms that using the mean efficiency for correction can produce inaccurate results and the effect of efficiency fluctuations is not negligible. Fortunately, the deviation is not negligible only when the efficiency difference \(\Delta\epsilon{}\) become unrealistically large compare to real experiments. In the case where the \(\Delta\epsilon{}\) is less than 0.2, the efficiency-corrected cumulants coincide with the input results within statistical uncertainties.

\begin{figure*}
\centering
\includegraphics[width=0.9\textwidth]{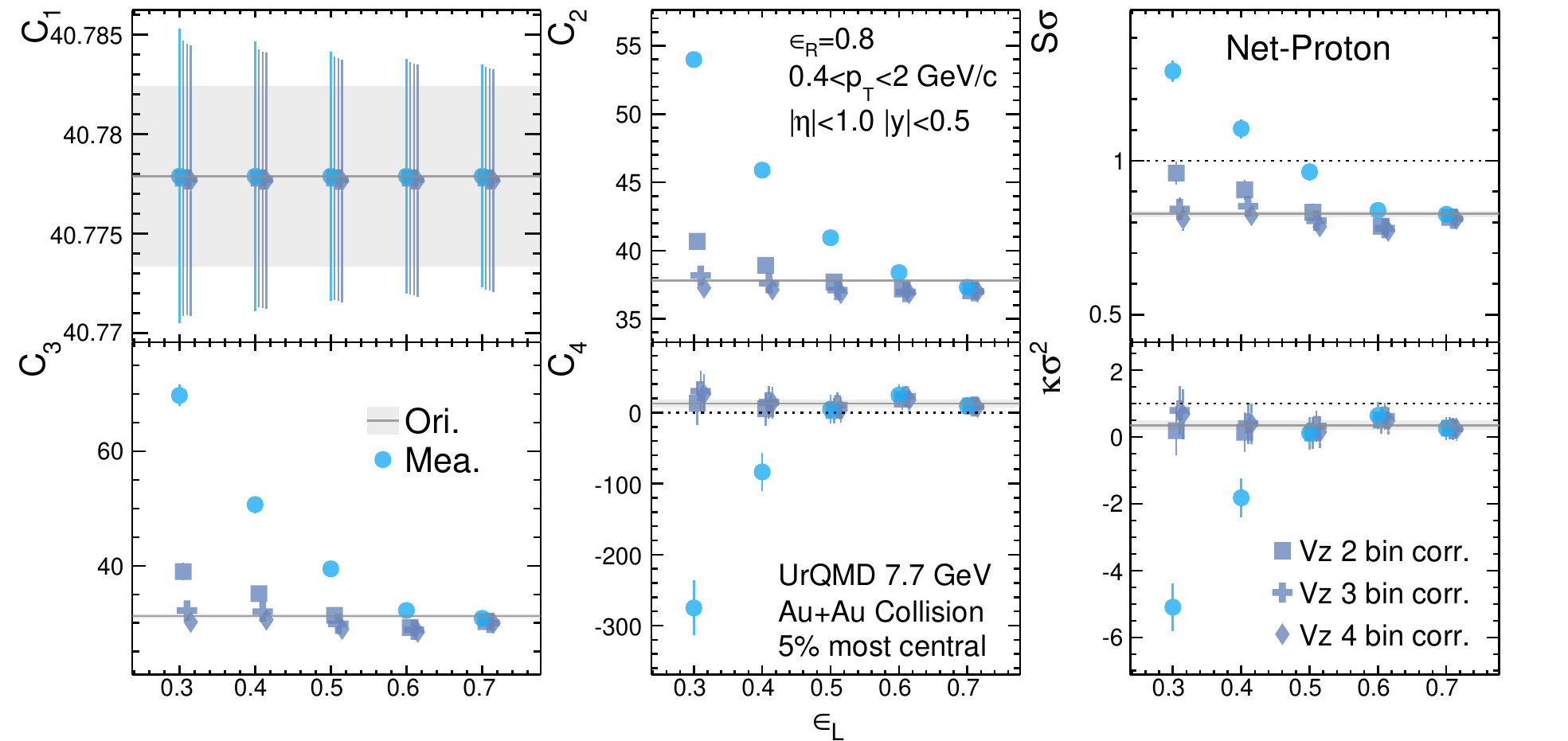}
\figcaption{\label{fig:urqmd-cumulants-to-delta-epsilon} (color online) The efficiency-corrected net-proton cumulants
with/without \(V_\mathrm{z}\) bin average in various
\(\Delta\epsilon{}\). The \(x\)-axis is the efficiency
\(\epsilon_\mathrm{L}\) in the left chamber, and the
\(\epsilon_\mathrm{R}\) is fixed at 0.8. The result without \(V_\mathrm{z}\) bin average (full circle marker) show large deviations from original results.}
\end{figure*}

\begin{figure*}
\centering
\includegraphics[width=0.9\textwidth]{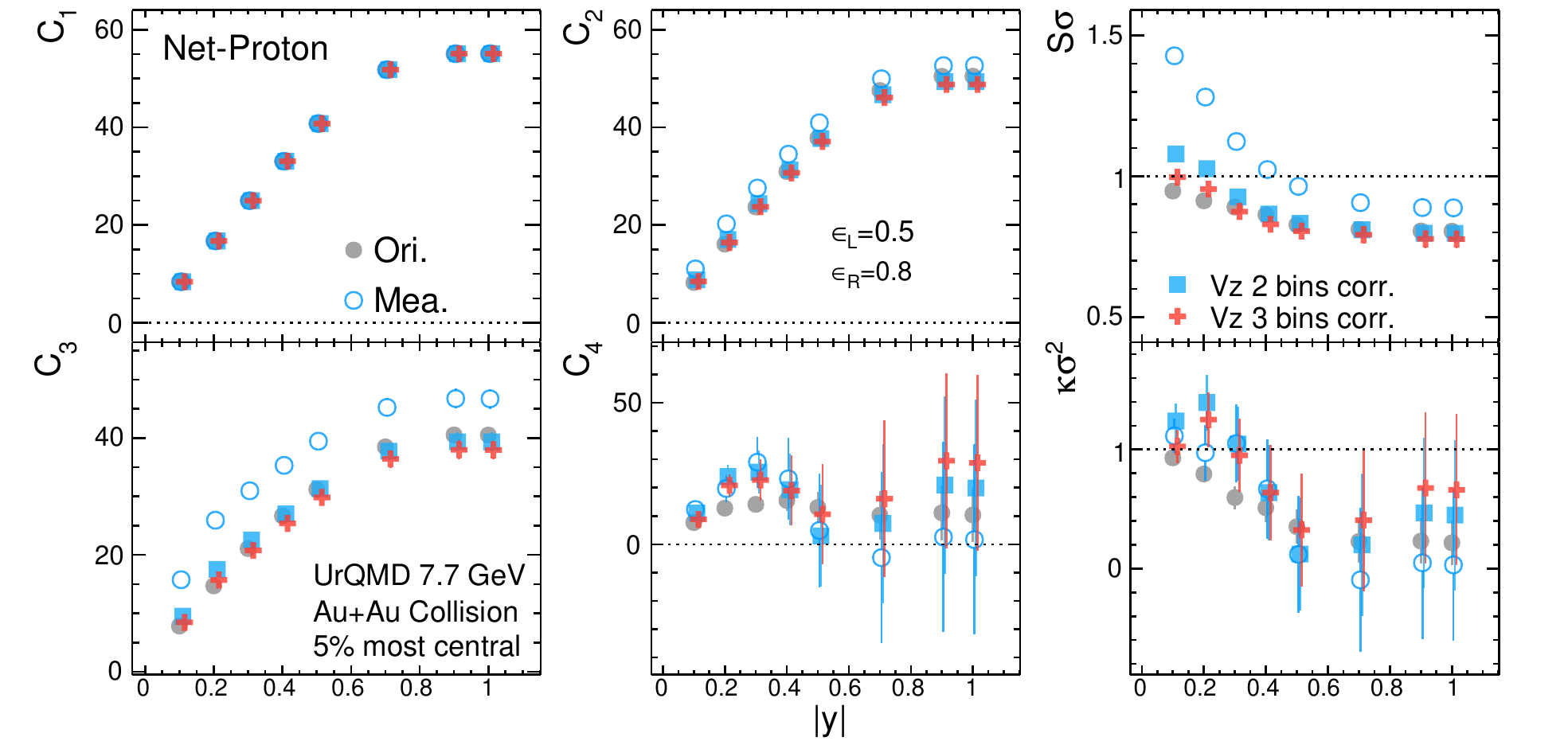}
\figcaption{\label{fig:urqmd-vz-corrected-cumulants-to-dy}(color online) Rapidity dependence of efficiency-corrected net-proton cumulants with/without \(V_\mathrm{z}\) bin average. The precision of correction has been significantly improved by an average of 2 \(V_\mathrm{z}\) bins. When the \(\Delta\epsilon{} = \epsilon_\mathrm{R} - \epsilon_\mathrm{L}\) is larger, finer \(V_\mathrm{z}\) bins (3 bins) are required.}
\end{figure*}

\subsection{$V_\text{z}$ bin correction}\label{vz-bin-correction}

The results shown in Fig.~\ref{fig:dN/deta} suggest that wider \(V_\mathrm{z}\) bins mix up more distinctive events, thus we can perform the efficiency correction within smaller \(V_\mathrm{z}\) bins.

The method to perform the correction of \(V_\mathrm{z}\) fluctuation is analogous to what was done in Section~{\ref{problem-of-using-mean-efficiency}}. We may suppose that events with different \(V_\mathrm{z}\) have their own mean efficiency. Therefore, the measured distribution is the superposed distribution from all \(V_\mathrm{z}\) ranges, and the superposed distribution from events with different efficiencies. Mean efficiencies of smaller \(V_\mathrm{z}\) bins can be obtained using Monte Carlo embedding procedures. Suppose the efficiency-corrected factorial moments at \(V_\mathrm{z}\) is \(F_{r}\left(V_\mathrm{z} \right)\), we can write the average result as equation~\eqref{eq:efficiency-bin-average-for-corrected-factorial-moments}
\[
    {\tilde{F}}_{r} = \int_{V_\mathrm{z}}^{}{F_{r}\left( V_\mathrm{z} \right)\; \mathrm{d}V_\mathrm{z}}.
\]
The statistical error is evaluated by propagation of the standard error.

We first investigate the effect of \(V_\mathrm{z}\) bin correction with different \(\Delta\epsilon{}\) and different \(V_\mathrm{z}\) bin width. Fig.~\ref{fig:urqmd-cumulants-to-delta-epsilon} shows that when \(\Delta\epsilon{}\) = \(\epsilon_\mathrm{R} - \epsilon_\mathrm{L}\) is large, the corrected results using the mean efficiency show large deviations from the input results. For the \(V_\mathrm{z}\) bin correction, we set up 2 or 3, 4 \(V_\mathrm{z}\) bins with equal interval in the range of (-50 cm, 50 cm) to perform efficiency correction. We found that the \(V_\mathrm{z}\) bin correction can significantly improve the results of high-order cumulants, especially in the case where \(\Delta\epsilon{}\) is large. Finer \(V_\mathrm{z}\) bin is important when \(\Delta\epsilon{}\) is large.  We also learn from Fig.~\ref{fig:urqmd-cumulants-to-delta-epsilon} that efficiency corrected 2 \(V_\mathrm{z}\) bins show great improvement. However, the effect of \(V_\mathrm{z}\) bin correction may depend on the rapidity acceptance, since we found in Fig.~\ref{fig:vz-dy-dependent-efficiency} that a smaller rapidity acceptance \(\Delta y\) exhibits a steeper slope on the \(V_\mathrm{z}\) dependence of efficiency. In Fig.~\ref{fig:urqmd-vz-corrected-cumulants-to-dy} we show 2 and 3 \(V_\mathrm{z}\) bins corrected results within various \(y\) cuts. When \(|y|\) cut is small, the result corrected by 3 bins is better than 2 bins. However, the discrepancy is not significant for larger \(|y|\) cut. The reason is that when \(|y|\) cut is large; particles are dispersed more evenly into two drifting chambers, so the efficiency depends less on \(V_\mathrm{z}\).

\section{Summary}\label{summary}

In heavy ion collision, the cumulants of event-by-event multiplicity of conserved charge have been used as sensitive observables to probe the QCD phase transition and the critical point. Since the experiments have a finite acceptance and detector efficiency, the measured distribution should be corrected for detecting efficiency in subsequent analysis.

The particle efficiencies of events measured under different experimental conditions should be different. We called this effect event-by-event efficiency fluctuations. The final event-by-event multiplicity distributions should be the superposed distributions of various type of events measured under different conditions. However, the efficiency correction is performed using the mean efficiency of the event sample. Mean efficiencies obtained from Monte Carlo embedding procedures have been used to do the efficiency correction for cumulants. The mean efficiencies reflect the net contribution of the acceptance, tracking efficiency and other effects. We have shown the relation between the factorial moments of the superposed distribution and the factorial moments from individual distributions (i.e.,\ the distribution with different efficiencies). We determined that the superposed factorial moments are the weighted average of the individual factorial moments, and the mean efficiency cannot restore the original input since the efficiency fluctuates across the various distributions. So we suggest that one should be very careful when binning the events into various sub-event samples, in which the efficiency variation is relatively small.

We have done a numerical simulation which combined two types of events with different efficiencies which revealed that a correction that is implemented using the mean efficiency can have a significant deviation from the original input. In addition, a more concrete simulation with the UrQMD model indicates that similar effects can occur in real experiments. In the UrQMD simulation, we consider the event-by-event fluctuation of \(z\)-coordination of collision vertex (\(V_\mathrm{z}\)). We also introduced the different working conditions of detectors at the west and east endcap of the detector (TPC), which can lead to unequal detecting efficiencies of the tracks in the west and east subpart of the chamber. We show that the event-by-event efficiency fluctuation effects can cause the efficiency-corrected cumulants using the mean efficiencies, to deviate from the original input. However, when the efficiency fluctuation is at the level of real experiments which is much smaller than our settings, the deviation can be neglected. We also attempted to reduce the efficiency variation by introducing the \(V_\mathrm{z}\) bin average method, which can significantly improve the precision of the efficiency correction.

The event-by-event efficiency fluctuation implemented in our simulation, is not the only source that can be present in real experiments. For example, other sources may exist such as the variation of the detector performance as a function of time and bad events outliers in the multiplicity distributions. To obtain high precise and reliable efficiency-corrected cumulants, it is necessary to careful study event selection and classification to ensure the event-by-event efficiency fluctuations is small. This work presents a simple but effective method to improve the precision of efficiency correction for cumulant analysis in relativistic heavy-ion collision experiments.

\end{multicols}

\vspace{10mm}

\section*{Appendices A}\label{appendix}

\begin{multicols}{2}
\small
\setcounter{equation}{0}
\renewcommand{\theequation}{A\arabic{equation}}

\subsection*{Efficiency correction for factorial moments}\label{efficiency-correction-for-factorial-moments}

The factorial moments generating function is \begin{equation}\label{eq:factorial-moment-generating-function} G_{F}\left( s \right) = \sum_{n = 0}^{\infty}{P\left( n \right)s^{n}} \end{equation} where the \(n\) is the value of the stochastic variable and the \(P\left( n \right)\) is the probability density function. The summation can also be expressed as:
\[
    G_{F}\left( s \right) = \left\langle s^{n} \right\rangle
\]
Therefore the \(r\)-th factorial moment of \(n\) is given as:
\begin{align}
    F_{r} & = \left. \frac{\partial^{r}}{\partial s^{r}}G_{F}\left( s \right) \right|_{s = 1} \nonumber \\
          & = \sum_{n = r}^{\infty}{P\left( n \right) n\left( n - 1 \right)\left( n - 2 \right)\cdots\left( n - r + 1 \right)}\nonumber  \\
          & = \left\langle n\left( n - 1 \right)\cdots\left( n - r + 1 \right) \right\rangle
\end{align}
The probability of detecting \(n\) particles (\(p\left( n \right)\)) is given by the Binomial distribution \(B_{N}\left( n,\ \epsilon \right)\), where \(N\) is the number of input particles and the \(\epsilon{}\) is the efficiency.
\begin{align}\label{eq:detecting-probability}
    p\left( n \right) &= \sum_{N = n}^{\infty}{P\left( N \right)B_{N}\left( n,\epsilon \right)} \nonumber \\
                      &= \sum_{N = n}^{\infty}{P\left( N \right)\binom{N}{n}\epsilon^{n}{\left( 1 - \epsilon \right)}^{N - n}}
\end{align}
The generating function of the measured factorial moments is then
\begin{align}\label{eq:measured-factorial-moments-generating-function}
G_{f}\left( s \right) & = \sum_{n = 0}^{\infty}{p\left( n \right)s^{n}} \nonumber \\
                      & = \sum_{n = 0}^{\infty}\sum_{N = n}^{\infty}{P\left( N \right)\binom{N}{n}\epsilon^{n}{\left( 1 - \epsilon \right)}^{N - n}}s^{n} \nonumber \\
                      & = \sum_{N = 0}^{\infty}{P\left( N \right)}\sum_{n = 0}^{N}{\binom{N}{n}\epsilon^{n}{\left( 1 - \epsilon \right)}^{N - n}}s^{n} \nonumber \\
                      & = \sum_{N = 0}^{\infty}{P\left( N \right)}\sum_{n = 0}^{N}{\binom{N}{n}{\left( {\epsilon s} \right)}^{n}{\left( 1 - \epsilon \right)}^{N - n}}
\end{align}
The last line can be simplified using the Binomial theorem to give:
\[
    \sum_{n = 0}^{N}{\binom{N}{n} {\left( {\epsilon s} \right)}^{n} {\left( 1 - \epsilon \right)}^{N - n}} = \left\lbrack \epsilon s + \left( 1 - \epsilon \right) \right\rbrack^{N}
\]
We then have
\begin{align}\label{eq:ori-gen-func-to-mea-gen-func}
    G_{f}\left( s \right) & = \sum_{N = 0}^{\infty}{P\left( N \right)\left\lbrack 1 + \epsilon\left( s - 1 \right) \right\rbrack^{N}} \nonumber \\
                          & = \sum_{N = 0}^{\infty}{P\left( N \right){{s'}^{N}} = G_{F}(s')}
\end{align}
where \(s' = \left[ 1 + \epsilon\left( s - 1 \right) \right]{}\). We then have the relation between the measured factorial moments \(f_{r}\) and the original factorial moments \(F_{r}\)
\begin{align}\label{eq:measured-from-original-factorial-moments}
    f_{r} & = \left.\frac{\partial^{r}}{\partial s^{r}}G_{f}\left( s \right) \right|_{s = 1} \nonumber \\
          & = \left.\frac{\partial^{r}}{\partial s^{r}}\sum_{N = 0}^{\infty}{P\left( N \right){\left[ 1 + \epsilon\left( s - 1 \right) \right]}^{N}} \right|_{s = 1} \nonumber \\
          & = \sum_{N = r}^{\infty}{\epsilon^{r}P\left( N \right)N\left( N - 1 \right)\cdots\left( N - r + 1 \right)} = \epsilon^{r}F_{r}
\end{align}

\subsection*{Multivariate factorial moments}\label{multivariate-factorial-moments}

In the report, we use multivariate factorial moments to describe the net-proton number. The net-proton factorial moments has 2 dimensions which describe the proton and anti-proton number respectively. The generating function of \(q\)-dimensional factorial moments is an extension to equation~\eqref{eq:factorial-moment-generating-function}
\begin{equation}\label{nd-factorial-moments-generating-function}
    G_{\boldsymbol{F}}(\boldsymbol{t})=\prod_{q} \sum_{n_q=0}^{\infty}{P_q(n_q)t_q^{n_q}}
\end{equation}
Therefore
\begin{align}
    \boldsymbol{F}_{\boldsymbol{r}} & = \prod_{q}{\left.\frac{\partial^{r_q}}{\partial t_q^{r_q}} \sum_{n_q=0}^{\infty}{P_q(n_q)t_q^{n_q}}\right|_{t_q = 1}} \nonumber \\
                                    & = \left\langle \prod_{q}{{\left(t_q\right)}_{r_q}} \right\rangle
\end{align}
where the symbol \({\left(t_q\right)}_{r_q}\) is a falling factorial
\[
    {\left(t_q\right)}_{r_q} = t_q (t_q-1) (t_q-2) \cdots (t_q-r_q+1)
\]
The detecting probability density function for each kind of particle is identical to equation~\eqref{eq:detecting-probability}
\begin{equation}
    p_q(n_q) = \sum_{N_q=n_q}^{\infty} P_q(N_q)B_{N_q}(n_q, \epsilon_q)
\end{equation}
Therefore, the generating function for measured factorial moments is given by equation~\eqref{eq:measured-factorial-moments-generating-function}
\begin{align}\label{nd-measured-factorial-moments}
    G_{\boldsymbol{f}}(\boldsymbol{t}) & = \prod_{q} \sum_{n_q=0}^{\infty} p_q(n_q) t_q^{n_q} \nonumber \\
                                       & = \prod_{q} \sum_{n_q=0}^{\infty}\sum_{N_q=n_q}^{\infty} P_q(N_q)B_{N_q}(n_q, \epsilon_q) \nonumber \\
                                       & = \prod_{q} P_q(N_q){\left[ 1 + \epsilon_q(t_q - 1)\right]}^{N_q - n_q}
\end{align}
Thus, the efficiency correction relation is similar to equation~\eqref{eq:measured-from-original-factorial-moments}
\begin{equation}\label{eq:nd-measured-from-original-factorial-moments}
    f_{\boldsymbol{r}} = \left(\prod_{q}\epsilon_{q}^{r_q}\right)F_{\boldsymbol{r}}
\end{equation}
The conversation from \(q\)-dimensional factorial moments to \(q\)-dimensional moments is similar to equation~\eqref{eq:factorial-moments-to-moments}
\begin{equation}\label{eq:nd-factorial-moments-to-moments}
    \left\langle \prod_{q} N_q^{r_q}\right\rangle = \sum_{i_1=0}^{r_1} \cdots \sum_{i_q=0}^{r_q}s_2(r_1, i_1) \cdots s_2(r_q, i_q)F_{r_1,r_2,\ldots,r_q}
\end{equation}
With \(q\)-dimensional moments, we can write down the moments of any combination of \(q\) kinds of particles, for example, the moments of net-proton number is given as:
\begin{align}\label{eq:net-proton-moments}
    \left\langle{\left(N_1 - N_2\right)}^{r}\right\rangle & = \left\langle \sum_{i=0}^r{\binom{r}{i} {(-1)}^i N_1^{r-i}N_2^{i}} \right\rangle \nonumber \\
                                                          & = \sum_{i=0}^{r}{\binom{r}{i}} {{(-1)}^i}\left\langle N_1^{r-i}N_2^{i}\right\rangle\nonumber \\
                                                          & = \sum_{i=0}^{r}{\binom{r}{i}} {{(-1)}^i}\sum_{k_1=0}^{r-i}\sum_{k_2=0}^{i}s_2(r-i, k_1)s_2(i, k_2)F_{k_1k_2}
\end{align}
The cumulants of the net-proton number is given by equation~\eqref{eq:moments-to-cumulants} directly.

\end{multicols}

\vspace{-2.5mm} \centerline{\rule{80mm}{0.1pt}} \vspace{1mm}

\begin{multicols}{2}

\bibliographystyle{unsrt}
\bibliography{ref}

\begin{thebibliography}{10}

\bibitem{Gupta:2011wh}
Sourendu Gupta, Xiaofeng Luo, Bedangadas Mohanty, Hans~Georg Ritter, and Nu~Xu.
\newblock {Scale for the Phase Diagram of Quantum Chromodynamics}.
\newblock {\em Science}, 332:1525--1528, 2011.

\bibitem{Cleymans:1990nz}
J.~Cleymans, H.~Satz, E.~Suhonen, and D.~W. von Oertzen.
\newblock {Strangeness Production in Heavy Ion Collisions at Finite Baryon
  Number Density}.
\newblock {\em Phys. Lett.}, B242:111--114, 1990.

\bibitem{Davidson:1991um}
N.~J. Davidson, H.~G. Miller, R.~M. Quick, and J.~Cleymans.
\newblock {Chemical equilibration in heavy ion collisions}.
\newblock {\em Phys. Lett.}, B255:105--109, 1991.

\bibitem{Lao:2017skd}
Hai-Ling Lao, Fu-Hu Liu, Bao-Chun Li, and Mai-Ying Duan.
\newblock {Kinetic freeze-out temperatures in central and peripheral
  collisions: Which one is larger?}
\newblock {\em Nucl. Sci. Tech.}, 29:82, 2018.

\bibitem{Adamczyk:2014ipa}
L.~Adamczyk et~al.
\newblock {Beam-Energy Dependence of the Directed Flow of Protons, Antiprotons,
  and Pions in Au+Au Collisions}.
\newblock {\em Phys. Rev. Lett.}, 112(16):162301, 2014.

\bibitem{Brachmann:1999xt}
J.~Brachmann, S.~Soff, A.~Dumitru, Horst Stoecker, J.~A. Maruhn, W.~Greiner,
  L.~V. Bravina, and D.~H. Rischke.
\newblock {Antiflow of nucleons at the softest point of the EoS}.
\newblock {\em Phys. Rev.}, C61:024909, 2000.

\bibitem{Song:2017wtw}
Huichao Song, You Zhou, and Katarina Gajdosova.
\newblock {Collective flow and hydrodynamics in large and small systems at the
  LHC}.
\newblock {\em Nucl. Sci. Tech.}, 28(7):99, 2017.

\bibitem{Hattori:2016emy}
Koichi Hattori and Xu-Guang Huang.
\newblock {Novel quantum phenomena induced by strong magnetic fields in
  heavy-ion collisions}.
\newblock {\em Nucl. Sci. Tech.}, 28(2):26, 2017.

\bibitem{Aoki:2006we}
Y.~Aoki, G.~Endrodi, Z.~Fodor, S.~D. Katz, and K.~K. Szabo.
\newblock {The Order of the quantum chromodynamics transition predicted by the
  standard model of particle physics}.
\newblock {\em Nature}, 443:675--678, 2006.

\bibitem{Schaefer:2011ex}
B.~J. Schaefer and M.~Wagner.
\newblock {QCD critical region and higher moments for three flavor models}.
\newblock {\em Phys. Rev.}, D85:034027, 2012.

\bibitem{Asakawa:2009aj}
Masayuki Asakawa, Shinji Ejiri, and Masakiyo Kitazawa.
\newblock {Third moments of conserved charges as probes of QCD phase
  structure}.
\newblock {\em Phys. Rev. Lett.}, 103:262301, 2009.

\bibitem{Endrodi:2011gv}
G.~Endrodi, Z.~Fodor, S.~D. Katz, and K.~K. Szabo.
\newblock {The QCD phase diagram at nonzero quark density}.
\newblock {\em JHEP}, 04:001, 2011.

\bibitem{Stephanov:2004wx}
Mikhail~A. Stephanov.
\newblock {QCD phase diagram and the critical point}.
\newblock {\em Prog. Theor. Phys. Suppl.}, 153:139--156, 2004.
\newblock [Int. J. Mod. Phys.A20,4387(2005)].

\bibitem{Chen:2015dra}
Jiunn-Wei Chen, Jian Deng, Hiroaki Kohyama, and Lance Labun.
\newblock {Robust characteristics of nongaussian fluctuations from the NJL
  model}.
\newblock {\em Phys. Rev.}, D93(3):034037, 2016.

\bibitem{Vovchenko:2015pya}
V.~Vovchenko, D.~V. Anchishkin, M.~I. Gorenstein, and R.~V. Poberezhnyuk.
\newblock {Scaled variance, skewness, and kurtosis near the critical point of
  nuclear matter}.
\newblock {\em Phys. Rev.}, C92(5):054901, 2015.

\bibitem{Vovchenko:2016rkn}
Volodymyr Vovchenko, Mark~I. Gorenstein, and Horst Stoecker.
\newblock {van der Waals Interactions in Hadron Resonance Gas: From Nuclear
  Matter to Lattice QCD}.
\newblock {\em Phys. Rev. Lett.}, 118(18):182301, 2017.

\bibitem{Fan:2017kym}
Wenkai Fan, Xiaofeng Luo, and Hongshi Zong.
\newblock {Identifying the presence of the critical end point in QCD phase
  diagram by higher order susceptibilities}, 2017.

\bibitem{Fukushima:2014lfa}
Kenji Fukushima.
\newblock {Hadron resonance gas and mean-field nuclear matter for baryon number
  fluctuations}.
\newblock {\em Phys. Rev.}, C91(4):044910, 2015.

\bibitem{Karsch:2010ck}
Frithjof Karsch and Krzysztof Redlich.
\newblock {Probing freeze-out conditions in heavy ion collisions with moments
  of charge fluctuations}.
\newblock {\em Phys. Lett.}, B695:136--142, 2011.

\bibitem{BraunMunzinger:2011dn}
P.~Braun-Munzinger, B.~Friman, F.~Karsch, K.~Redlich, and V.~Skokov.
\newblock {Net-proton probability distribution in heavy ion collisions}.
\newblock {\em Phys. Rev.}, C84:064911, 2011.

\bibitem{Gavai:2010zn}
R.~V. Gavai and Sourendu Gupta.
\newblock {Lattice QCD predictions for shapes of event distributions along the
  freezeout curve in heavy-ion collisions}.
\newblock {\em Phys. Lett.}, B696:459--463, 2011.

\bibitem{Borsanyi:2013hza}
S.~Borsanyi, Z.~Fodor, S.~D. Katz, S.~Krieg, C.~Ratti, and K.~K. Szabo.
\newblock {Freeze-out parameters: lattice meets experiment}.
\newblock {\em Phys. Rev. Lett.}, 111:062005, 2013.

\bibitem{Bazavov:2012vg}
A.~Bazavov et~al.
\newblock {Freeze-out Conditions in Heavy Ion Collisions from QCD
  Thermodynamics}.
\newblock {\em Phys. Rev. Lett.}, 109:192302, 2012.

\bibitem{Morita:2014fda}
Kenji Morita, Bengt Friman, and Krzysztof Redlich.
\newblock {Criticality of the net-baryon number probability distribution at
  finite density}.
\newblock {\em Phys. Lett.}, B741:178--183, 2015.

\bibitem{Jiang:2015cnt}
Lijia Jiang, Pengfei Li, and Huichao Song.
\newblock {Multiplicity fluctuations of net protons on the hydrodynamic
  freeze-out surface}.
\newblock {\em Nucl. Phys.}, A956:360--364, 2016.

\bibitem{Jiang:2015hri}
Lijia Jiang, Pengfei Li, and Huichao Song.
\newblock {Correlated fluctuations near the QCD critical point}.
\newblock {\em Phys. Rev.}, C94(2):024918, 2016.

\bibitem{Adamczyk:2014fia}
L.~Adamczyk et~al.
\newblock {Beam energy dependence of moments of the net-charge multiplicity
  distributions in Au+Au collisions at RHIC}.
\newblock {\em Phys. Rev. Lett.}, 113:092301, 2014.

\bibitem{Alba:2014eba}
Paolo Alba, Wanda Alberico, Rene Bellwied, Marcus Bluhm, Valentina
  Mantovani~Sarti, Marlene Nahrgang, and Claudia Ratti.
\newblock {Freeze-out conditions from net-proton and net-charge fluctuations at
  RHIC}.
\newblock {\em Phys. Lett.}, B738:305--310, 2014.

\bibitem{Luo:2017faz}
Xiaofeng Luo and Nu~Xu.
\newblock {Search for the QCD Critical Point with Fluctuations of Conserved
  Quantities in Relativistic Heavy-Ion Collisions at RHIC : An Overview}.
\newblock {\em Nucl. Sci. Tech.}, 28(8):112, 2017.

\bibitem{Ejiri:2005wq}
S.~Ejiri, F.~Karsch, and K.~Redlich.
\newblock {Hadronic fluctuations at the QCD phase transition}.
\newblock {\em Phys. Lett.}, B633:275--282, 2006.

\bibitem{Stephanov:1999zu}
Misha~A. Stephanov, K.~Rajagopal, and Edward~V. Shuryak.
\newblock {Event-by-event fluctuations in heavy ion collisions and the QCD
  critical point}.
\newblock {\em Phys. Rev.}, D60:114028, 1999.

\bibitem{Stephanov:2008qz}
M.~A. Stephanov.
\newblock {Non-Gaussian fluctuations near the QCD critical point}.
\newblock {\em Phys. Rev. Lett.}, 102:032301, 2009.

\bibitem{Stephanov:2011pb}
M.~A. Stephanov.
\newblock {On the sign of kurtosis near the QCD critical point}.
\newblock {\em Phys. Rev. Lett.}, 107:052301, 2011.

\bibitem{Bzdak:2016sxg}
Adam Bzdak, Volker Koch, and Nils Strodthoff.
\newblock {Cumulants and correlation functions versus the QCD phase diagram}.
\newblock {\em Phys. Rev.}, C95(5):054906, 2017.

\bibitem{Luo:2012kja}
Xiaofeng Luo.
\newblock {Search for the QCD Critical Point by Higher Moments of Net-proton
  Multiplicity Distributions at STAR}.
\newblock {\em Nucl. Phys.}, A904-905:911c--914c, 2013.
\newblock [Central Eur. J. Phys.10,1372(2012)].

\bibitem{Friman:2014cua}
Bengt Friman.
\newblock {Probing the QCD phase diagram with fluctuations}.
\newblock {\em Nucl. Phys.}, A928:198--208, 2014.

\bibitem{Kitazawa:2011wh}
Masakiyo Kitazawa and Masayuki Asakawa.
\newblock {Revealing baryon number fluctuations from proton number fluctuations
  in relativistic heavy ion collisions}.
\newblock {\em Phys. Rev.}, C85:021901, 2012.

\bibitem{Kitazawa:2012at}
Masakiyo Kitazawa and Masayuki Asakawa.
\newblock {Relation between baryon number fluctuations and experimentally
  observed proton number fluctuations in relativistic heavy ion collisions}.
\newblock {\em Phys. Rev.}, C86:024904, 2012.
\newblock [Erratum: Phys. Rev.C86,069902(2012)].

\bibitem{Friman:2011pf}
B.~Friman, F.~Karsch, K.~Redlich, and V.~Skokov.
\newblock {Fluctuations as probe of the QCD phase transition and freeze-out in
  heavy ion collisions at LHC and RHIC}.
\newblock {\em Eur. Phys. J.}, C71:1694, 2011.

\bibitem{Cheng:2008zh}
M.~Cheng et~al.
\newblock {Baryon Number, Strangeness and Electric Charge Fluctuations in QCD
  at High Temperature}.
\newblock {\em Phys. Rev.}, D79:074505, 2009.

\bibitem{Luo:2015doi}
Xiaofeng Luo.
\newblock {Exploring the QCD Phase Structure with Beam Energy Scan in Heavy-ion
  Collisions}.
\newblock {\em Nucl. Phys.}, A956:75--82, 2016.

\bibitem{Aggarwal:2010cw}
M.~M. Aggarwal et~al.
\newblock {An Experimental Exploration of the QCD Phase Diagram: The Search for
  the Critical Point and the Onset of De-confinement}, 2010.

\bibitem{Zhao:2016djo}
Ameng Zhao, Xiaofeng Luo, and Hongshi Zong.
\newblock {Baryon Number Fluctuations in Quasi-particle Model}.
\newblock {\em Eur. Phys. J.}, C77(4):207, 2017.

\bibitem{Jeon:1999gr}
S.~Jeon and V.~Koch.
\newblock {Fluctuations of particle ratios and the abundance of hadronic
  resonances}.
\newblock {\em Phys. Rev. Lett.}, 83:5435--5438, 1999.

\bibitem{Asakawa:2000wh}
Masayuki Asakawa, Ulrich~W. Heinz, and Berndt Muller.
\newblock {Fluctuation probes of quark deconfinement}.
\newblock {\em Phys. Rev. Lett.}, 85:2072--2075, 2000.

\bibitem{Xu:2016mqs}
Ji~Xu.
\newblock {Energy Dependence of Moments of Net-Proton, Net-Kaon, and Net-Charge
  Multiplicity Distributions at STAR}.
\newblock {\em J. Phys. Conf. Ser.}, 736(1):012002, 2016.

\bibitem{Xu:2016qjd}
Ji~Xu, Shili Yu, Feng Liu, and Xiaofeng Luo.
\newblock {Cumulants of net-proton, net-kaon, and net-charge multiplicity
  distributions in Au + Au collisions at $\sqrt {s_{NN}}$=7.7 , 11.5, 19.6, 27,
  39, 62.4, and 200 GeV within the UrQMD model}.
\newblock {\em Phys. Rev.}, C94(2):024901, 2016.

\bibitem{Adamczyk:2017wsl}
L.~Adamczyk et~al.
\newblock {Collision Energy Dependence of Moments of Net-Kaon Multiplicity
  Distributions at RHIC}, 2017.

\bibitem{Aggarwal:2010wy}
M.~M. Aggarwal et~al.
\newblock {Higher Moments of Net-proton Multiplicity Distributions at RHIC}.
\newblock {\em Phys. Rev. Lett.}, 105:022302, 2010.

\bibitem{Adamczyk:2013dal}
L.~Adamczyk et~al.
\newblock {Energy Dependence of Moments of Net-proton Multiplicity
  Distributions at RHIC}.
\newblock {\em Phys. Rev. Lett.}, 112:032302, 2014.

\bibitem{Luo:2015ewa}
Xiaofeng Luo.
\newblock {Energy Dependence of Moments of Net-Proton and Net-Charge
  Multiplicity Distributions at STAR}.
\newblock {\em PoS}, CPOD2014:019, 2015.

\bibitem{Mukherjee:2016kyu}
Swagato Mukherjee, Raju Venugopalan, and Yi~Yin.
\newblock {Universal off-equilibrium scaling of critical cumulants in the QCD
  phase diagram}.
\newblock {\em Phys. Rev. Lett.}, 117(22):222301, 2016.

\bibitem{Mukherjee:2015swa}
Swagato Mukherjee, Raju Venugopalan, and Yi~Yin.
\newblock {Real time evolution of non-Gaussian cumulants in the QCD critical
  regime}.
\newblock {\em Phys. Rev.}, C92(3):034912, 2015.

\bibitem{Nahrgang:2014fza}
Marlene Nahrgang, Marcus Bluhm, Paolo Alba, Rene Bellwied, and Claudia Ratti.
\newblock {Impact of resonance regeneration and decay on the net-proton
  fluctuations in a hadron resonance gas}.
\newblock {\em Eur. Phys. J.}, C75(12):573, 2015.

\bibitem{Bzdak:2012an}
Adam Bzdak, Volker Koch, and Vladimir Skokov.
\newblock {Baryon number conservation and the cumulants of the net proton
  distribution}.
\newblock {\em Phys. Rev.}, C87(1):014901, 2013.

\bibitem{Ling:2015yau}
Bo~Ling and Mikhail~A. Stephanov.
\newblock {Acceptance dependence of fluctuation measures near the QCD critical
  point}.
\newblock {\em Phys. Rev.}, C93(3):034915, 2016.

\bibitem{Berdnikov:1999ph}
Boris Berdnikov and Krishna Rajagopal.
\newblock {Slowing out-of-equilibrium near the QCD critical point}.
\newblock {\em Phys. Rev.}, D61:105017, 2000.

\bibitem{He:2016uei}
Shu He, Xiaofeng Luo, Yasushi Nara, ShinIchi Esumi, and Nu~Xu.
\newblock {Effects of Nuclear Potential on the Cumulants of Net-Proton and
  Net-Baryon Multiplicity Distributions in Au+Au Collisions at
  $\sqrt{s_{\text{NN}}} = 5\,\text{GeV}$}.
\newblock {\em Phys. Lett.}, B762:296--300, 2016.

\bibitem{Schuster:2009jv}
Marlene Nahrgang, Tim Schuster, Michael Mitrovski, Reinhard Stock, and Marcus
  Bleicher.
\newblock {Net-baryon-, net-proton-, and net-charge kurtosis in heavy-ion
  collisions within a relativistic transport approach}.
\newblock {\em Eur. Phys. J.}, C72:2143, 2012.

\bibitem{Sakaida:2014pya}
Miki Sakaida, Masayuki Asakawa, and Masakiyo Kitazawa.
\newblock {Effects of global charge conservation on time evolution of cumulants
  of conserved charges in relativistic heavy ion collisions}.
\newblock {\em Phys. Rev.}, C90(6):064911, 2014.

\bibitem{Abelev:2008ab}
B.~I. Abelev et~al.
\newblock {Systematic Measurements of Identified Particle Spectra in $p p, d^+$
  Au and Au+Au Collisions from STAR}.
\newblock {\em Phys. Rev.}, C79:034909, 2009.

\bibitem{Bzdak:2013pha}
Adam Bzdak and Volker Koch.
\newblock {Local Efficiency Corrections to Higher Order Cumulants}.
\newblock {\em Phys. Rev.}, C91(2):027901, 2015.

\bibitem{Luo:2014rea}
Xiaofeng Luo.
\newblock {Unified description of efficiency correction and error estimation
  for moments of conserved quantities in heavy-ion collisions}.
\newblock {\em Phys. Rev.}, C91(3):034907, 2015.

\bibitem{Nonaka:2017kko}
Toshihiro Nonaka, Masakiyo Kitazawa, and ShinIchi Esumi.
\newblock {More efficient formulas for efficiency correction of cumulants and
  effect of using averaged efficiency}.
\newblock {\em Phys. Rev.}, C95(6):064912, 2017.

\bibitem{Kitazawa:2017ljq}
Masakiyo Kitazawa and Xiaofeng Luo.
\newblock {Properties and uses of factorial cumulants in relativistic heavy-ion
  collisions}.
\newblock {\em Phys. Rev.}, C96(2):024910, 2017.

\bibitem{Luo:2011tp}
Xiaofeng Luo.
\newblock {Error Estimation for Moments Analysis in Heavy Ion Collision
  Experiment}.
\newblock {\em J. Phys.}, G39:025008, 2012.

\bibitem{Luo:2013bmi}
Xiaofeng Luo, Ji~Xu, Bedangadas Mohanty, and Nu~Xu.
\newblock {Volume fluctuation and auto-correlation effects in the moment
  analysis of net-proton multiplicity distributions in heavy-ion collisions}.
\newblock {\em J. Phys.}, G40:105104, 2013.

\bibitem{Leo:1987kd}
W.~R. Leo.
\newblock {\em {Techniques for Nuclear and Particle Physics Experiments: A How
  to Approach}}.
\newblock Berlin, Germany: Springer (1987) 368 p, 1987.

\bibitem{Bass:1998ca}
S.~A. Bass et~al.
\newblock {Microscopic models for ultrarelativistic heavy ion collisions}.
\newblock {\em Prog. Part. Nucl. Phys.}, 41:255--369, 1998.
\newblock [Prog. Part. Nucl. Phys.41,225(1998)].

\bibitem{He:2017zpg}
Shu He and Xiaofeng Luo.
\newblock {Proton Cumulants and Correlation Functions in Au + Au Collisions at
  $\sqrt{s_\mathrm{NN}}$=7.7-200 GeV from UrQMD Model}.
\newblock {\em Phys. Lett.}, B774:623--629, 2017.

\end{thebibliography}

\end{multicols}
\clearpage
\end{document}